\documentclass[10pt]{iopart}

\usepackage[english]{babel}
\usepackage[colorlinks,citecolor=red,urlcolor=blue,bookmarks=false,hypertexnames=true]{hyperref}
\usepackage{amsthm}
\usepackage{cite}
\usepackage{iopams}
\usepackage{graphicx}

\expandafter\let\csname equation*\endcsname\relax
\expandafter\let\csname endequation*\endcsname\relax
\expandafter\let\csname subequations*\endcsname\relax
\expandafter\let\csname endsubequations*\endcsname\relax

\usepackage{amsmath, amssymb, esint}
\usepackage{cases}
\usepackage{accents}
\usepackage{mathtools}
\usepackage{tensor}
\usepackage[math]{blindtext}
\usepackage[margin=2.5cm]{geometry}
\usepackage{stackrel}
\usepackage{slashed}
\usepackage{citesort}
\usepackage{graphicx}
\usepackage{float}

\def\divider{\par
  \vskip 0.5em
  \centerline{\hbox to 0.5\hsize{\hrulefill}}
  \vskip 0.5em
}

\begin{document}

\title{Liouville theory on a horizon: point particle/scalar field duality and Page-like curve}

\author{J-B. Roux$^\ast$}

\address{$^\ast$ Université d'Aix-Marseille, Marseille, \textsc{France} (Now unaffiliated, independent researcher)}

\ead{jeanbaptiste.roux@live.fr}
\begin{minipage}{\textwidth}
\begin{abstract}
    We show that the consequences of a recent paper on quantum gravity are 1) a duality between point particles and massive scalar propagators, 2) the recovery of the entropy of a boundary (a black hole) in the same form as that of the EFT approach to Quantum Gravity and 3) a quantum correction to Hawking radiations and a Page-like curve. In this recent paper, information about what lies inside a boundary is encoded onto it, meaning that in this approach the information directly leaks from the horizon to the bulk in the form of Hawking radiations.\\
    \noindent{\textsc{keywords}--- \it{Quantum Gravity, Black hole entropy}\,\text{|}}
    \submitto{\CQG}
\end{abstract}
\maketitle
\end{minipage}

\section{Introduction}\label{s1}

Mathematically speaking, black holes are thought to be the key to properly understanding quantum gravity, because of the so-called singularity beneath the horizon. It is commonly believed that such singularity indicates that our current understanding of gravitational physics breaks down and that a quantum theory of gravity should supersede General Relativity at such scales. However, one should be careful in what one calls a ``black hole''. The first non-trivial solution to Einstein's field equations was the Schwarzschild metric, describing the exterior of a non-rotating, spherically symmetric star. Later, Oppenheimer and Snyder developed their eponymous model of gravitational collapse, first criticized for its oversimplifications, but later accepted as relevant in astrophysics because it matched well the more elaborated simulations of star collapse. The Oppenheimer-Snyder model precludes that geodesics end at a singularity. It uses a cycloidal time $\tau$, that is, a time parameter whose range is finite and goes from $\tau=0$ to $\tau=\pi$, where $\tau=\pi$ is the final state of the model. However, for the exterior observer, this collapse takes an infinite time, and for the falling observer, we do not even know if the universe still exists such as we know it, at $\tau=\pi$. Thus, astrophysical, actual, black holes may not exist in the form of a Schwarzschild or Kerr solution, which both preclude that the final state of a collapsing star already exists at $t\rightarrow -\infty$.\\
\indent Having this in mind, one could question the relevance of the information paradox. Indeed, if a Schwarzschild or Kerr black hole evaporates by emitting particles carrying no information about the interior of the horizon, information seems lost at the end of the evaporation. But with unphysical assumptions, it is common to stumble upon unphysical conclusions, such as here: loss of unitarity in a theory that should be so. Maybe, all the attempts to solve the information paradox are doomed to be consequences of mathematical artifacts, because the very black holes we study are mathematical artifacts too. Indeed, actual black holes are more likely ever-contracting matter than a Schwarzschild solution already existing at $t\rightarrow -\infty$. Still, studying black holes as mathematical models is interesting because of their simplicity. \\
\indent In a recent paper \cite{Roux:2024bna}, it has been shown that following a certain approach to quantum gravity, the partition function of General Relativity with cosmological constant reduces to two Liouville models. This approach consists of describing a hypersurface as a point in the super-space and then using the standard quantum mechanical argument to find the propagator of a point particle in curved space. The partition function is defined with fixed boundary conditions and the Liouville models are defined on it. One can view this two-dimensional surface as a fixed boundary. Specifically, the partition function of Einstein's gravity is entirely described by the 3D gravity theory on the initial (at $t=0$) hypersurface, itself entirely described by its boundary. Because it is supposed to be common to all hypersurfaces, the cosmological horizon seems to be a good choice for such a boundary. Nonetheless, it is possible to choose a fixed boundary different from this horizon and calculate relevant quantities such as correlation functions of given operators.\\
\indent In this article, we show a duality between point particles and fields, in the sense that the expectation values of a point source and a scalar propagator are identical. Then, we use this property to describe with fields (rather than point particles) how the information enclosed by the boundary lies on it. In doing so, we obtain the quantum-corrected entropy of a black hole. Moreover, we obtain an entropy following a Page-like curve (as it is unclear whether this is truly the Page curve or not), along with a quantum correction to Hawking radiations.\\
\indent The article is organized as follows: In part \ref{s2}, we rewrite some results of \cite{Roux:2024bna} and adapt them to the presence of fields with sources. In part \ref{s3}, we calculate the expectation value of a point-like source and a scalar propagator, where we stumble upon our duality. The same is done for a vector propagator in the \ref{appendix}. Then, in part \ref{s4}, we rewrite the entropy of the theory by splitting the partition function into a classical part, and a quantum perturbation part. Upon introducing on the boundary a pure state of information of, for example, point particles, we show that the quantum correction to the classical Bekenstein-Hawking entropy modifies the entropy. Furthermore, this entropy gives all the corrections of perturbative quantum gravity \textit{a la} Barvinsky-Vilkovisky with fixed coefficients. This means the theory given in \cite{Roux:2024bna} is predictive. In part \ref{s5} we demonstrate how information can escape the horizon, which sources the scalar field at the cosmological boundary. We also show how to recover a Page-like curve by calculating independently how entropy evolves in our model, and how it evolves in QFT in curved space (once we embed this QFT onto a codimension 1 hypersurface, as explained at the end of \ref{s2}).

\section{Preliminary results}\label{s2}

In \cite{Roux:2024bna}, it was shown that the partition function of Quantum Gravity (with a cosmological constant) with the Keldysh contour as time integration domain leads to two $H_3^+$-WZW models. This is an instance of the $\text{AdS}_3/\text{CFT}_2$ duality. The cosmological constant is positive but in the process of dimensional reduction of the theory on a codimension 1 hypersurface at $t=0$, the signature of the metric is such that the theory is as if the cosmological constant was negative. In \cite{Roux:2024bna} the boundary taken was the cosmological boundary, but due to the BRST treatment of General Relativity, the lapse function $N$ is changed by a gauge fixing parameter $f^H$ \cite{Alexandrov_1998}. Since it is a free function, we can choose it to have whatever form we want. Specifically, we can choose it to be zero, and hence define a boundary, elsewhere than the cosmological boundary. This point is crucial because the theory is agnostic to at least one point inside a closed boundary. So changing the $f^H=0$ surface allows us to cover all the possible points in the bulk of the codimension 1 hypersurface at $t=0$. Using the gauge fields $A^+$ and $A^-$ constructed from the spatial spin connections and the triads, one obtains the partition function of General Relativity with Keldysh contour:
\begin{equation}\label{eq6}
    Z_\text{K.}[A^\pm|_\partial,0] = \left\langle e^{\frac{i}{2\pi} \int d^2z \text{tr}[A_z^\pm|_\partial \overline{\mathcal{J}}{}^\pm]}\right\rangle_{\text{WZW}^\pm_{\frac{c_A}{2}\pm k}}.
\end{equation}
Where $\overline{\mathcal{J}}$ is the anti-holomorphic current of the model, $c_A$ the quadratic Casimir invariant of the group $\text{SL}(2,\mathbb{C})/\text{SU}(2) \equiv H_3^+$, and $k = \frac{\beta_m}{\sqrt{\Lambda}}$ is the level of the theory \cite{Yildirim_2015} (with $2\beta_m$ the minimal time-lapse of the theory). This last quantity shows that for the level to be an integer, we ought to have $\beta_m = n \sqrt{\Lambda},\,\,\,n\in \mathbb{N}\setminus\{0\}$ to have large gauge transformation invariance, and in our case, $\beta_m$ is related to a minimum time lapse in the theory. The Wess-Zumino-Witten (WZW) action is as follows.
\begin{equation}\label{eq6.1}
    S_\mathrm{WZW}[g]= \frac{1}{2\pi}\int_{\partial \Sigma} \mathrm{tr}[\partial_z g \wedge \partial_{\overline{z}}g^{-1}]+\frac{i}{12\pi} \int_\mathbb{B} \mathrm{tr}\left[ (g^{-1} d g)^{\wedge 3}\right].
\end{equation}
In this action, $g$ is a group-valued field on a closed two-dimensional surface $\partial\Sigma$, seen as the boundary of $\mathbb{B}$. The first term is the kinetic one, while the second is topological and has the property to be independent of the geometry of $\mathbb{B}$ (this is why we did not choose to write $\Sigma$ instead of $\mathbb{B}$). It is important for the level $k$ of the theory to be an integer because otherwise, the path integral $\int \mathcal{D}g\,e^{kS_\mathrm{WZW}[g]}$ is not invariant by large gauge transformations. The identification (\ref{eq6}) comes from either a full treatment of a Chern-Simons path integral to express it as a WZW model, or upon noticing that the constraint equation (found by integrating the Lagrange multiplier $A_r$) on the Chern-Simons path integral with fixed boundary:
\begin{equation}\label{eq6.2}
    F_{z\overline{z}} Z_\mathrm{CS}[A^\pm |_\partial]=0,
\end{equation}
is the same as the equation governing (\ref{eq6}). Indeed, it suffices to express the functional operator $F_{z\overline{z}}$ of gauge curvature with $A^\pm_z$ and $A^\pm_{\overline{z}}\propto \frac{\delta}{\delta A^\pm_z}$. Now, as done in \cite{Roux:2024bna}, we can further reduce (\ref{eq6}) to a Liouville theory upon choosing:
\begin{equation}\label{eq7}
    A^{(3)}_z|_\partial(z) = \frac{2}{\rho}a^{(3)}_z\sum_{i=1}^N \alpha_i\left[\frac{\rho}{z-z_i}+\frac{1}{2i\pi} \oint_{\partial D(z)} dw \frac{1}{w-z_i}\right].
\end{equation}
With $a^{(3)}_z = \sqrt{\sigma}^{-1}$, the inverse of the square root of the 2D Fubini-Study metric determinant, and $\rho = 2\beta_m \sqrt{\sigma}^{-1/2}$ the radius of the disc $D(z)$. Furthermore, $z_i$ are evenly scattered positions of the punctures onto the boundary, as depicted by the divergences of (\ref{eq7}). This choice does not change the entropy. But the partition function is more conveniently written as follows, after a Gauss decomposition of $g$ in (\ref{eq6.1}):
\begin{equation}\label{eq8}
    Z_\text{K.}[\alpha_i|z_i] = \left[ \left\langle \prod_{i=1}^{A/8} e^{2\alpha_i \varphi(z_i)}e^{2\alpha^\ast_i \varphi(z_i)} \right\rangle_\text{L.} \right]^2.
\end{equation}
Where $\alpha_i = 1+\frac{i}{2}P_i$, for $P_i$ a momentum located at $z_i$, and $A/8$ is half of the number of area units ($4\ell_\text{Pl}^2$) composing the boundary, and $\varphi(z)$ is the field involved in the Gauss decomposition of $g$ (specifically, this is the field in the exponential of the diagonal element in the basis of the group $H_3^+$). The (regularized) Liouville action in the expectation value is written as \cite{Li_2020} (we have the asymptotic $\varphi \stackrel{z\rightarrow z_i}{\sim} -2\alpha_i \ln(|z-z_i|)$ and $\varphi \stackrel{z\rightarrow \infty}{\sim} -2\ln|z|$):
\begin{equation}\label{eq8.1}
    S_\mathrm{L.}[\varphi]=\frac{1}{\pi} \int_{\partial \Sigma} d^2z \sqrt{\sigma}\left( \sigma^{\mu \nu}\partial_\mu \varphi \partial_\nu \varphi+ QR\varphi+\frac{1}{4}e^{2b\varphi} \right)+\varphi_\infty+2\ln(r_\infty)-\sum_i(\alpha_i \varphi_i+2\alpha_i^2 \ln(\epsilon_i)).
\end{equation}
With $\varphi_\infty = \frac{1}{2\pi r_\infty}\int_{\partial D_\infty}dw\,\varphi$ and $\varphi_i = \frac{1}{2\pi \epsilon_i}\int_{\partial D(z_i)}dw\,\varphi$ (where $r_\infty$ is the radius of $\partial D_\infty$ and $\epsilon_i$ the radius of $\partial D(z_i)$). Moreover, $b=1$, $Q=b+\frac{1}{b}=2$ is related to the central charge $c$ of the theory by the formula $c=1+6Q^2$, and $R$ is the two-dimensional Ricci scalar of $\partial \Sigma$. Of course, in this article all the expressions involving $e^{2\alpha \varphi(z)}$ are not normal-ordered and $\varphi(z)$ is not circle-averaged. These procedures are to be performed to do rigorously the calculations \cite{chatterjee2024liouvilletheoryintroductionrigorous}. But since we will deal with the classical approximation to the partition function only we can get rid of these manipulations. The classical approximation is found by splitting the integration measure $\mathcal{D}g \leadsto \mathcal{D}(g_\mathrm{cl.} \mathfrak{g})$ in (\ref{eq6}), and using the Polyakov-Wiegmann identity:
\begin{align}\label{eq8.2}
    Z[A_z|_\partial] =& \int \mathcal{D}g\,e^{\left( k\pm\frac{c_A}{2} \right)S_\text{WZW}[g]+\frac{ik}{2\pi}\int_\partial d^2z\,\text{tr}[A_z|_\partial (g^{-1}\partial_{\overline{z}}g)]}
\nonumber \\
    \leadsto& \int \mathcal{D}(g_\text{cl.} \mathfrak{g})\,e^{\left( k\pm\frac{c_A}{2} \right)\left[S_\text{WZW}[g_\text{cl.}]+S_\text{WZW}[\mathfrak{g}]-\frac{i}{2\pi}\int d^2z \,\text{tr}[g_\text{cl.}^{-1} \partial_z g_\text{cl.}\,\mathfrak{g}^{-1}\partial_{\overline{z}} \mathfrak{g}]\right]}e^{\frac{ik}{2\pi}\int_\partial d^2z\,\text{tr}[A_z|_\partial (g_\text{cl.}^{-1}\partial_{\overline{z}}g_\text{cl.}+\mathfrak{g}^{-1} \partial_{\overline{z}} \mathfrak{g})]}
\nonumber \\
    \leadsto& \det(g_\text{cl.})e^{\left( k\pm\frac{c_A}{2} \right)S_\text{WZW}[g_\text{cl.}]+\frac{ik}{2\pi}\int_\partial d^2z\,\text{tr}[A_z|_\partial (g_\text{cl.}^{-1}\partial_{\overline{z}}g_\text{cl.})]} \int \mathcal{D}\mathfrak{g}\,e^{\left( k\pm\frac{c_A}{2} \right)S_\text{WZW}[\mathfrak{g}]+\frac{ik}{4\pi}\int_\partial d^2z\,\text{tr}[A_z|_\partial (\mathfrak{g}^{-1} \partial_{\overline{z}} \mathfrak{g})]}.
\end{align}
The classical equation of motion of $g_\mathrm{cl.}$ is used for the last step. We will systematically recall this splitting when using it to find a classical approximation. The idea is to see that since there is a fixed background, usually a black hole, there is a fixed area parameter $A$ which depends on the inverse temperature $\beta$ of the black hole. This means the path integral in the last line of (\ref{eq8.2}) is a Liouville expectation value with a fixed area, while the prefactor is the exponential of the classical action. The latter coincides with $e^{-\frac{\beta^2}{16\pi}+\mathcal{O}(\Lambda)}$ (see \cite{Roux:2024bna}), as in four-dimensional Euclidean Quantum Gravity.\\
\indent To introduce matter in the theory, it suffices to add the potential of the matter theory (the gradient potential included) to the Hamiltonian of General Relativity. This results in an Euclidean 3D theory of matter onto the hypersurface at $t=0$. Thus, we can write formally the partition function divided by $Z^{\Phi,\text{3D}}[0]$ (the partition function of the 3D matter theory) as:
\begin{equation}\label{eq4}
    \widetilde{Z}_\text{K.}[A^\pm|_\partial,J] = \left\langle e^{\frac{4i\beta_m^2}{2}\int d^3x \sqrt{\gamma(x)} \int d^3y \sqrt{\gamma(y)} J(x) G(x,y) J(y)} \right\rangle_{\text{CS}^\pm}.
\end{equation}
With $\gamma$ the determinant of the induced metric on the codimension 1 hypersurface at $t=0$, and $G(x,y)$ the 3D propagator of the field. The idea is then to foresee that because the theory reduces to two WZW models, one should in principle determine the contribution of the scalar propagator through a functional derivative on $A^\pm_r$. Indeed, we can think of $\widetilde{Z}_\text{K.}[A^\pm|_\partial,0]$ as a wave functional constrained by the equation $F^\pm_{z\overline{z}}\widetilde{Z}_\text{K.}[A^\pm|_\partial,0]=0$, with $F^\pm_{z\overline{z}}$ the curvature of the gauge field $A^\pm_z$, as explained before. Then, the contributions of $G(x,y)$ should be dealt with via the standard method for introducing sources in a WZW model \cite{gawedzki1999conformalfieldtheorycase}. Specifically, we are interested in the following quantity:
\begin{equation}\label{eq5}
    \frac{1}{4\beta_m^2}\left.\frac{-i\delta}{\delta J(x)}\frac{-i\delta}{\delta J(y)}\widetilde{Z}_\text{K.}[A^\pm|_\partial,J]\right|_{J=0}.
\end{equation}
Using the standard Worldline Formalism (WF) to express $G(x,y)$ as an integral of an exponential, we can deal with this quantity that resembles the insertion of generalized Wilson lines in its form. This means we can express (\ref{eq4}) as a $H_3^+$-WZW expectation value

\section{Introducing matter}\label{s3}

\subsection{Free scalar theory}\label{ss3.1}

In this part, and the following one, we force the gauge fixing parameter $f^H$ to coincide with the lapse function of the Schwarzschild metric in Schwarzschild coordinates. Since we have a candidate theory of Quantum Gravity, it would be interesting to see what happens when we add matter. To begin, we introduce point particles and then use a scalar field.
For point matter, we need to use the following source, which is nothing but the exponential of a relativistic point particle action ($M$ is the mass of the particle):
\begin{align}\label{eq9}
    e^{iM\int_0^{2\beta_m}dt \sqrt{\gamma_{ij} \dot{x}^i \dot{x}^j}} = e^{i\frac{M}{2\sqrt{\Lambda}}\int_0^{2\beta_m}dt \sqrt{\frac{1}{2}\text{tr}[(\sigma^a A^a_i \dot{x}^i)^2]}}.
\end{align}
Because the gauge field $A_r$ in the Chern-Simons formulation of 3D gravity is a Lagrange multiplier, we need to take the derivative on $A_r$ to find the wave-functional, which is our partition function. But our derivative is:
\begin{equation}\label{eq10}
    \frac{\delta}{\delta A_r}\left(i\frac{M}{2\sqrt{\Lambda}}\int_0^{2\beta_m}dt \sqrt{\frac{1}{2}\text{tr}[(A_i \dot{x}^i)^2]}\right)=\frac{iM}{2\sqrt{\Lambda}}\int_0^{2\beta_m}dt \frac{\dot{x}^r \dot{x}^iA_i}{\sqrt{\frac{1}{2}\text{tr}[(A_i \dot{x}^i)^2]}}\delta^{(2)}(z-w)\delta(r-r').
\end{equation}
We impose that $A^{(\overline{z})}_i x^i=0$, implying $x^{\overline{z}}=0$, so that the trace is $\text{tr}[(A_i \dot{x}^i)^2] = 2 (A^{(3)}_i \dot{x}{}^i)^2$. This is true for the $H_3^+$ basis that may be built from the Pauli matrices. We can see that taking this assumption $A^{(\overline{z})}_i x^i=0$ into account, we obtain the expectation value of a Wilson line. Taking the normalized trace of the Wilson line we obtain:
\begin{align}\label{eq11}
    & \frac{1}{2}\text{Tr}\left[e^{i\frac{M}{2\sqrt{\Lambda}}\int_0^{2\beta_m}dt \dot{x}^r \sigma^{(3)} \int_w^zdy\,A_z}\right]
\nonumber \\
    =& \frac{1}{2}\text{Tr}\left[e^{iM \dot{x}^r \sigma^{(3)} \int_w^zdy\,A_z}\right]
\nonumber \\
    =& e^{iM \dot{x}^r (\varphi(z)-\varphi(w))}.
\end{align}
Where we used the expression of $A_z$, which is $A_z = g^{-1}\partial_z g$, and injected the Gauss decomposition in it. Introducing, in the same manner, $\mathcal{N}$ point particles we obtain the sourced partition function:
\begin{align}\label{eq12}
    Z_\text{K.}[\alpha_i | z_i | \mathcal{N}] \stackrel{!}{=}&  \left[ \left\langle \prod_{n=1}^{\mathcal{N}}\left(e^{2\alpha_n\varphi(z_{i_n})}e^{2 \alpha^\ast_n\varphi(z_{j_n})}\right)\prod_{k\notin \{i\},\{j\}} e^{2\alpha_k \varphi(z_k)} \right\rangle_\text{L.} \right]^2.
\end{align}
With $\alpha_n = \frac{Q}{2} + \frac{i}{2}M_n\dot{x}_n^r$ ($\{M_n\}_n$ is a set of masses associated with the point particles). As we can see, the ``momenta'' of the vertex operators are truly momenta. We can interpret this as particles leaving the points $z_{j_n}$, and going towards $z_{i_n}$.\\
\indent Now, we introduce fields to see if this result holds still. \textit{A priori}, there is no reason for fields to give the same expression than that of point particles. But to introduce matter realistically, point particles are not the adapted objects to use, while fields are. We are thus interested in the expression:
\begin{equation}\label{eq13}
    \langle G(x,y) \rangle_{\text{CS}^\pm} = \left\langle \frac{1}{\Delta^\text{3D}-m^2-\frac{1}{4} R}\delta(x,y) \right\rangle_{\text{CS}^\pm}.
\end{equation}
Where $\Delta^\mathrm{3D}$ is the three-dimensional (3D) Laplacian in curved space, $m$ the mass of the field, and $R$ the 3D Ricci scalar. Moreover, the Dirac delta is in curved space. To evaluate this expression we use the standard procedure of the Worldline Formalism \cite{corradini2021spinning}. This is achieved by using the Schwinger parametrization of the propagator and using a suitable integration measure, namely $\mathcal{D}x \sqrt{\gamma(x)}$. Then, to be able to do the calculation, one expresses the square root of the metric determinant as a path integral over Lee-Yang ghosts $\mathfrak{a}$, $\mathfrak{b}$ and $\mathfrak{c}$ (with $\mathfrak{a}$ a bosonic variable and the others are fermionic.) As we will see, the $\frac{1}{4}R$ term will be exactly compensated by the standard counter-term used in dimensional regularization. To see this, we momentarily change $\frac{1}{4}R \leadsto \xi R$ in the propagator (\ref{eq13}). Thus, we write for a Euclidean Schwinger parameter $s$:
\begin{align}\label{eq14}
    G(x,y) =& \frac{1}{\Delta^\text{3D}-m^2-\xi R}\delta(x,y) = \int_0^\infty ds\, e^{-sm^2} \int_{x(0)=x}^{x(s)=y} \mathcal{D}x\,\sqrt{\gamma} e^{-\int_0^s d\sigma \left( \frac{1}{4}\gamma_{ij}\dot{x}{}^i \dot{x}{}^j+ \left(\xi-\frac{1}{4}\right) R\right)}
\nonumber \\
    =& \int_0^\infty ds\, e^{-sm^2} \int_{x(0)=x}^{x(s)=y} \mathcal{D}x \int \mathcal{D}\mathfrak{a}\mathcal{D}\mathfrak{b}\mathcal{D}\mathfrak{c}\, e^{-\int_0^s d\sigma \left( \frac{1}{4}\gamma_{ij}(\dot{x}{}^i \dot{x}{}^j+\mathfrak{a}^i\mathfrak{a}^j + \mathfrak{b}^i \mathfrak{c}^j) + \left(\xi-\frac{1}{4}\right) R\right)}.
\end{align}
Because the metric is roughly the square of our gauge fields, and because we seek a formulation that resembles a Wilson line, where the gauge field appears linearly, we use a Hubbard-Stratonovich transformation (which essentially amounts to the proportionality $e^{-a^2y^2} \propto \int dx\,e^{-\frac{1}{4}x^2+i a x}$). Decomposing $\gamma_{ij} = \frac{1}{8\Lambda} \text{tr}[A_i A_j]$ we obtain, when $\xi = \frac{1}{4}$:
\begin{align}\label{eq15}
    &\int_0^\infty ds\, e^{-sm^2} \int_{x(0)=x}^{x(s)=y} \mathcal{D}x \int \mathcal{D}\mathfrak{a}\mathcal{D}\mathfrak{b}\mathcal{D}\mathfrak{c}\, e^{-\int_0^s d\sigma \left( \frac{1}{16\Lambda}\left[A^a_i\dot{x}{}^i A^a_j\dot{x}{}^j+A^a_i\mathfrak{a}^i A^a_j\mathfrak{a}^j + \mathcal{B}^{\text{T},a}M \mathcal{B}^a\right] \right)},\,\,\, \mathcal{B}^a \equiv A^a_i \left( \begin{matrix}
        \mathfrak{b}^i \\ \mathfrak{c}^i
    \end{matrix} \right)\equiv A^a_i\mathcal{B}^i
\nonumber \\
    \propto& \int_0^\infty ds\, e^{-sm^2} \int_{x(0)=x}^{x(s)=y} \mathcal{D}x \int \mathcal{D}\mathfrak{a}\mathcal{D}\mathfrak{b}\mathcal{D}\mathfrak{c} \int \mathcal{D} X\mathcal{D}\alpha \mathcal{D}\beta \mathcal{D}\gamma\, e^{-\int_0^s d\sigma 4\Lambda\left[ X^2+\alpha^2+\beta^{\text{T},a} M^{-1} \gamma^a \right]}
\nonumber \\    
    &\times e^{-i\int_0^s d\sigma \left[ X^a A^a_i \dot{x}{}^i+\alpha^a A^a_i \mathfrak{a}^i+\beta^{\text{T},a} \mathcal{B}^a+\mathcal{B}^{\text{T}, a} \gamma^a\right]}.
\end{align}
Where the matrix $M$ mixes the Lee-Yang ghost doublet such that we recover (\ref{eq14}). To manipulate lighter expressions, we focus only on the relevant part for the expectation value, namely the second exponential. We will rewrite it as a function of $\text{tr}[A_i \sigma^a]\propto A^a_i$. Let us introduce $\Psi[A_z|_\partial]$, a wave-functional satisfying the following equation (the same as (\ref{eq6.2}) but with a source given by the second exponential in (\ref{eq15})):
\begin{align}\label{eq16}
    \left(\frac{k}{4\pi}F_{z\overline{z}}- \frac{i}{2} \int_0^s d\sigma\left( X^a \sigma^{\text{T},a} \dot{x}{}^r+\alpha^a \sigma^{\text{T},a} \mathfrak{a}^r + \beta^{\text{T},a}\sigma^{\text{T},a} \mathcal{B}^r+\sigma^{\text{T},a} \mathcal{B}^{\text{T},r} \gamma^a\right) \delta^{(3)}(x(\sigma)-x) \right)\Psi[A_z|_\partial] = 0.
\end{align}
Where we omitted all the tensor products, and with $\sigma^a$ the $H_3^+$ basis. This equation is found by taking the expectation value of (\ref{eq15}) and performing the integral on $A_r$. This gives the same result as if we had differentiated on $A_r$ in the integrand of the expectation value. As we can see, the solution to this equation will be matrix-valued. But the expectation value ought to be a scalar. Because we have $\text{tr}[A_i \sigma^a]\propto A^a_i$, we take the determinant of $\Psi$, which is an exponential, to use the identity $\det \circ \exp = \exp \circ \mathrm{tr}$. The standard procedure to find a solution to this equation (which is gauge-invariant) is to first use a gauge transformation of $\Psi$ with a parameter $u$. Then, we pose $A^u=0$ in the bulk, and integrate on $a_z \equiv u^{-1}\partial_z u$. See \cite{gawedzki1999conformalfieldtheorycase} for more details on this procedure. By doing all this, we obtain:
\begin{equation}\label{eq17}
    \Psi'[A_z|_\partial] = \det\left(\left\langle e^{-\frac{i}{2}\int_0^s d\sigma\,\int_{z(0)}^{z(s)}a_z(x^r(\sigma),w)dw \left[ X^a \sigma^{\text{T},a} \dot{x}{}^r+\alpha^a \sigma^{\text{T},a} \mathfrak{a}^r + \beta^{\text{T},a}\sigma^{\text{T},a} \mathcal{B}^r+\sigma^{\text{T},a} \mathcal{B}^{\text{T},r} \gamma^a \right]} e^{\frac{i}{2\pi} \int_\mathbb{C} d^2z \text{tr}[A_z|_\partial \overline{\mathcal{J}}]} \right\rangle_\text{WZW}\right).
\end{equation}
Next, we use a Gauss decomposition of $u$, and the prescription (\ref{eq7}) to obtain:
\begin{equation}\label{eq18}
    \Psi'[\alpha_i|x_i] \propto \left\langle e^{-i\int_0^s d\sigma\,[\varphi(z(s))-\varphi(z(0))]\left[ X^{(3)} \dot{x}{}^r+\alpha^{(3)} \mathfrak{a}^r + \beta^{\text{T},(3)} \mathcal{B}^r + \mathcal{B}^{\text{T},r} \gamma^{(3)} \right]} e^{2\alpha_i \varphi(x_i)} \right\rangle_\text{L.}.
\end{equation}
This wave-functional is to be inserted in (\ref{eq15}) in place of the second exponential. The proportionality constant depends on the Lee-Yang ghosts, so the Berezin integrals over $\mathfrak{b}^z$ and $\mathfrak{c}^z$ are not zero. Doing the $(X,\alpha,\beta,\gamma)$ and Lee-Yang ghosts integrals we finally find:
\begin{align}\label{eq19}
    \langle G(x,y) \rangle \leadsto & \int_0^\infty ds\, e^{-sm^2} \int_{x(0)=x}^{x(s)=y} \mathcal{D}x^r \sqrt{\det_{\sigma\in[0,s]} [\varphi(z(s))-\varphi(z(0))]^2}\,e^{-\int_0^s d\sigma \left( \frac{1}{16\Lambda}[\varphi(z(s))-\varphi(z(0))]^2\dot{x}{}^r\dot{x}{}^r\right)}
\nonumber \\
    \propto& \int_0^\infty ds\, \frac{e^{-sm^2}}{(4\pi s)^{1/2}} e^{-\frac{(x^r-y^r)^2}{16\Lambda s}[\varphi(z(s))-\varphi(z(0))]^2}
\nonumber \\
    \propto&\, e^{-\frac{m|x^r-y^r|}{2\sqrt{\Lambda}}|\varphi(z(\infty))-\varphi(z(0))|}.
\end{align}
With $z(\infty) = z_y$ and $z(0)=z_x$. As we can see, when switching to the Lorentzian signature, there seems to be a difference between (\ref{eq19}) and (\ref{eq11}), because in the former there are absolute values. But interestingly, the classical approximation does not depend on whether $P=\frac{m}{2\sqrt{\Lambda}}(x^r-y^r)$ ($=M\dot{x}{}^r$ in (\ref{eq11})) is positive or negative, so we can drop the absolute value on $\varphi(z_y)-\varphi(z_x)$. Thus, we obtain:
\begin{align}\label{eq20}
    \langle G(x,y) \rangle_\text{K.} \stackrel{!}{=}&  \left[ \left\langle e^{2\alpha_P\varphi(z_{x})}e^{2\alpha^\ast_P\varphi(z_{y})}\prod_{k\neq x,y} e^{2\alpha_k \varphi(z_k)} \right\rangle_\text{L.} \right]^2 = Z_\text{K.}[\alpha_i|z_i|1].
\end{align}
Furthermore, we can notice that it works for a product of propagators too, and even a convolution works because we can express the measure $\sqrt{\gamma}d^3x$ via Lee-Yang ghosts degrees of freedom. From the first line of (\ref{eq19}), it is clear that the contribution of such measure is $\propto |\varphi(z)|$. We thus have a duality between point particles and scalar fields, which we discuss in part \ref{s6}. Now that we have a free field, we would like to add interactions, because realistic matter interacts. We could have used a more involved model like Quantum Electrodynamics where there are fermions and a vector boson, but as we can see, the calculations are already heavy with a simple scalar field. So in the following, we use a self-interacting scalar field with quartic potential.

\subsection{Interacting scalar theory}\label{ss3.2}

In the previous subsection, we have given an expression of the expectation value of a scalar field propagator, which is of the same form as the expectation value of the insertion of a point particle in the theory. Because the method relies on (\ref{eq14}) and (\ref{eq16}), it is clear that we can give an expression to the expectation value of a product of propagators, too:
\begin{equation}\label{eq21}
    \left\langle \prod_{k=1}^{\mathcal{N}} G(x_k,y_k)\right\rangle_{\text{K.}} = Z_\text{K.}[\alpha_i|z_i|\mathcal{N}].
\end{equation}
This means the duality between point particles and propagators is complete because in \cite{Roux:2024bna}, the expectation of a product of insertion of point particles is found to be exactly of the same form. In fact, every contribution of the metric that can be expressed as an operation on an exponential of the gauge fields $A^\pm_i$ can be transcribed into an expression involving $\propto |\varphi(z)|$. So, a vertex of the following form:
\begin{equation}\label{eq22}
    \includegraphics[scale=0.35]{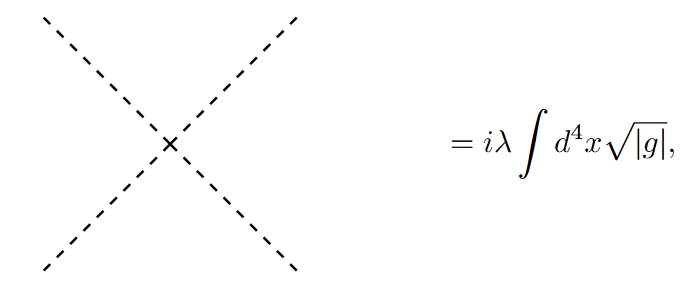}
\end{equation}
has to be expressed with the field $\varphi$. Specifically, we are searching for the expectation value of:
\begin{equation}\label{eq23}
    \left.\frac{1}{(2\beta_m)^4}\frac{\delta}{\delta J(x_1)}\frac{\delta}{\delta J(x_2)}\frac{\delta}{\delta J(x_3)}\frac{\delta}{\delta J(x_4)}e^{2i\beta_m\frac{\lambda}{4!} \int d^3x \sqrt{\gamma} \frac{1}{(2\beta_m)^4} \frac{\delta^4}{\delta J(x)^4}}e^{\frac{4i\beta_m^2}{2} \int d^3x \sqrt{\gamma(x)} \int d^3y \sqrt{\gamma(y)} J(x) G(x,y) J(y)} \right|_{J=0}.
\end{equation}
Notice that the very first term due to the expansion of the first exponential is $1$. Indeed, we divided by the free partition function with $J=0$. This can be seen as removing the vacuum bubbles. Without this removal of vacuum bubbles, we would have had to take the determinant of the free theory into account, expressing it using curvature invariants. As we will see later, this does not change much to \cite{Roux:2024bna}, because it essentially amounts to a renormalization of the different constants appearing in the 3D gravity model. The first term of (\ref{eq23}), which corresponds to (\ref{eq22}) with propagators, has the following expectation value (it is important for the Green's functions not to permute $x_i$ and $x$, as we are in 3D and not in space-time, so all the momenta should point toward the center of the vertex. For the case with more than one vertex treated in the subsection \ref{ss4.2}, we split the propagators between two vertices into two propagators.)
\begin{align}\label{eq24}
    & \left\langle i\lambda 2i\beta_m \int d^3x \sqrt{\gamma} \,G(x_1,x)G(x_2,x)G(x_3,x)G(x_4,x)\right\rangle_\text{K.}
\nonumber \\
    =& - \frac{\lambda}{2\beta_m} \int dr d^2 z\,r^2\left[\left\langle \varphi(z) \,\prod_{i=1}^4\left(e^{2\alpha_i \varphi(z_i)}e^{(\frac{Q}{4}-iP_i) \varphi(z)}\right)\prod_{k\notin \{i\}}e^{2\alpha_k \varphi(z_k)}\right\rangle_\text{L.}\right]^2.
\end{align}
Note that, because the integral is not an operation depending on the metric, we have to take it out of the square, and the same goes for the multiplicative factor $\lambda$. Because the horizon is discrete, the integral over $z$ becomes $\int d^2z \leadsto 4\ell_\text{Pl}^2\sum_k$. An important thing to note is that the ``vertex operators'' (the exponentials of $\varphi$) are not normal-ordered, because we focus on the classical approximation only. This means we can use the usual properties of the exponentials:
\begin{equation}\label{eq25}
    \varphi(z)\prod_{i=1}^4 e^{(\frac{Q}{4}-iP_i) \varphi(z)} \leadsto \left.\partial_\varepsilon e^{(Q+\varepsilon-i\sum_i P_i)\varphi(z)} \right|_{\varepsilon =0}.
\end{equation}
Thus, (\ref{eq24}) gives us the following expression, where we rearrange the vertex operators conveniently:
\begin{align}\label{eq26}
    (\ref{eq24}) \propto& -\frac{\lambda}{2\beta_m} \int dr\,r^2 \sum_l \left[\partial_\varepsilon \left.\left\langle \,e^{(Q+\varepsilon-i\sum_i P_i)\varphi(z_l)}\prod_{k}e^{2\alpha_k \varphi(z_k)}\right\rangle_\text{L.}\right|_{\varepsilon =0}\right]^2.
\end{align}
We perform an approximation consisting of neglecting the correlations between $\varphi(z_i)$, and keep the saddle point approximation only. The partition function with insertions is thus the free partition function multiplied by an exponential containing information of the insertions. However, the action must be regularized, as in \cite{Li_2020, Zamolodchikov_1996}, and $\varphi$ ought to have the right asymptotic expansion at infinity and near the insertions. We do not include regularization factors for the insertions because, conceptually, they do not belong to the free action. The counter-terms are of the form $\frac{1}{2}Q^2 \ln(|\eta|)$, with $|\eta|$ the distance between two points, which we force to be $\sim 2$ (the argument is the same as in \cite{Roux:2024bna}: the entropy is to be divided by an area of $4$ in Planck units). In doing so, we obtain that every contribution of the insertion amounts to $2^{-P_i^2}$ for $i=1,2,3,4$, and because $\sum_i k_i=0$ (where $k_i$ are the momenta for each insertion of (\ref{eq21})), we obtain:
\begin{align}\label{eq27}
    (\ref{eq24}) \sim& -\frac{\lambda}{2\beta_m} \int_0^{r_s} dr \,r^2\sum_l\left[ \partial_\varepsilon \left.2^{(\varepsilon-i\sum_i P_i)^2-\sum_i P_i^2}\right|_{\varepsilon =0}\right]^2 \left( \frac{A}{\pi} \right)^{-2Q/b}.
\nonumber \\
    \stackrel{\Lambda \rightarrow 0}{\rightarrow}& \frac{\lambda}{2\beta_m} \times \frac{-\ln(2)}{5} A \left( \frac{A}{\pi} \right)^{-2Q/b} 2^{-\frac{4m^2}{4\beta_m^2}\left(10r_s^2-5r_s \sum_i x_i^r+\frac{1}{2}((\sum_i x^r_i)^2+\sum_i {x^r_i}^2) \right)}\left( 4r_s-\sum_i x_i^r \right).
\end{align}
To obtain this expression, we used $P_i=\frac{m(x_i-r)}{2\beta_m}$. This term (\ref{eq27}) does not scale like the free contribution, which means it can become greater than it. This implies that interactions can become increasingly more important as the number of vertices grows. As we can see, (\ref{eq27}) is relevant only near the event horizon. In this part, we focused on the embedding of a (interacting) scalar field inside a black hole horizon. In the next part, we will derive the entropy of the system ``black hole$+$scalar field''.

\section{Quantum-corrected entropy}\label{s4}

\subsection{Free scalar field}\label{ss4.1}

Now that we have introduced matter and seen the possibility of encoding the scalar propagator onto the boundary, we would like to calculate the entropy of a black hole having matter enclosed by its horizon. Note that in the process of finding the partition function, one decomposes the integration measure $\mathcal{D}h$ of the $H_3^+$-WZW model (previously written $\mathcal{D}g$ in part \ref{s2}) into two parts $\mathcal{D}(h_\text{cl.} \mathfrak{h})$. So the metric stays still everywhere, while the gauge parameter we integrate on is changed. This means there's no need to decompose the metric as $\gamma=\gamma_\text{cl.}+\text{perturbations}$. This was to be expected because the metric is taken only on the boundary, which is fixed. We recall that from \cite{Roux:2024bna}, the expression of the partition function can be found to be (the Liouville model is now obviously with fixed area):
\begin{equation}\label{eq28}
    Z[\alpha_k|z_k] = e^{-\frac{\beta^2}{16\pi}+\mathcal{O}(\Lambda)}\left[\left\langle \prod_k e^{\alpha_k \varphi(z_k)}\right\rangle_\text{L.}\right]^2.
\end{equation}
With $\sum_{k}\Im(\alpha_k)=0$. Evaluating the expectation value by brute force is too difficult so we will rely on the same approximation as the section \ref{s3}. This approximation takes the classical approximation to the expectation value and neglects the position-dependent correlations. Let us suppose we have a side cut of the horizon. Then there are pairs of points on the boundary that are linked (that is, have the same value). We can thus represent this by the following situation:
\begin{equation}\label{eqchord0}
    \includegraphics[scale=1.25]{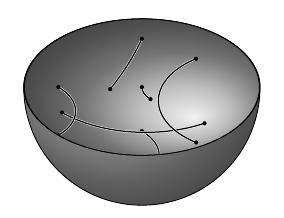}
\end{equation}
Thus, taking into account the boundary conditions, we have the leading contribution coming from the non-perturbative partition function with regularized action at infinity (but not at the insertions of momenta because conceptually, they do not belong to the total action; this means this time, the counterterm is of the form $Q^2 \ln(|\eta|)$, with $|\eta|=2$ as before, because $2\alpha_k$ is changed into $\alpha_k$ in the process of splitting the partition function measure) \cite{Li_2020,Giribet_2022,Zamolodchikov_1996}:
\begin{align}\label{eq30.2}
    Z[P|z] \sim& e^{-\frac{\beta^2}{16\pi}}\times \left( \frac{A}{\pi} \right)^{-2Q/b} \times 2^{-P^2}
\nonumber \\
    =& e^{-\frac{\beta^2}{16\pi}}\times \left( \frac{A}{\pi} \right)^{-4} \times 2^{-P^2}.
\end{align}
Indeed, $Q=2$ in our models. We denoted $P^2\equiv \sum_i p_i^2$. Thus, expressing $A$ as a function of the inverse temperature $\beta$ of the black hole, we find an entropy of the form:
\begin{equation}\label{eq31}
    S \sim \frac{A}{4}+8-4\ln\left(\frac{A}{\pi}\right) - \ln(2)P^2+\mathcal{O}(\Lambda).
\end{equation}
Until now the black hole is Euclidean, and at $t=0$. Now, we can let the black hole evolve through time and see what happens to its entropy. Let us suppose the black hole at $t=0$ is formed out of a pure state. This assumption enforces $S(0)=0$ and $S(t_\text{evap})=0$ (by unitarity), where $t_\text{evap}$ is the evaporation time of the black hole. The former enforces:
\begin{equation}\label{eq32}
     \frac{A_0}{4}+8-4\ln\left(\frac{A_0}{\pi}\right)=\ln(2){P^{(0)}}^2.
\end{equation}
The right-hand side can be interpreted as how the horizon geometry can source far-away radiations (this point will be clearer in part \ref{s5}). We are seeking an expression that is of the form (to recover the known result of the Effective Field Theory approach to Quantum Gravity \cite{Donoghue_1994}):
\begin{equation}\label{eq35}
    S=\frac{A}{4}+c_3(\mu)+\gamma \left[ \ln\left( \frac{A\mu^2}{4\pi} \right)+2-2\gamma_\text{E}\right].
\end{equation}
Notice that $\gamma = -4$. Thus, we obtain $8-8\ln(2)-8P^2 \ln(2)=c_3(\mu)-8+8\gamma_\text{E}-4\ln \left(\mu^2 \right)$. Note $\mu^2 \in [0,M^2_\text{Pl})$ (with $M^2_\text{Pl} = 1$ in Planck units). So, we find for the constant $c_3(\mu)$:
\begin{equation}\label{eq38}
    c_3(\mu,P) = 16-8\gamma_\text{E}-8\ln(2)-8P^2 \ln(2)+4\ln\left( \frac{\mu^2}{M^2_\text{Pl}} \right).
\end{equation}
This coefficient has the advantage of being of the form $c_3(\mu)=c_3(1)-\gamma \ln\left( \frac{\mu^2}{M^2_\text{Pl}} \right)$, which means that in principle, the entropy is invariant by renormalization group. Injecting (\ref{eq38}) into the entropy (\ref{eq35}), and assuming a direct correspondence between the method we used to find the entropy and Wald's method, we deduce that our theory has a low energy effective action:
\begin{align}\label{eq39}
    I[g] =& \int_\mathcal{M} d^4x \sqrt{-g}\left[ \frac{1}{2\kappa}R+\frac{1}{64\pi^2}\left( c_1(\mu) R^2+c_2(\mu) R_{\mu \nu}R^{\mu \nu}+c_3(\mu) R_{\mu \nu \alpha \beta} R^{\mu \nu \alpha \beta} \right)\right.
\nonumber \\
    & \left. -\frac{1}{64\pi^2}\left( \alpha R \ln \left( \frac{\square}{\mu^2} \right) R+\beta R_{\mu \nu} \ln \left( \frac{\square}{\mu^2} \right) R^{\mu \nu}+\gamma R_{\mu \nu \alpha \beta} \ln \left( \frac{\square}{\mu^2} \right) R^{\mu \nu \alpha \beta} \right)\right].
\end{align}
Where $\square = \nabla_\mu \nabla^\mu$ is the d'Alembertian in curved space. The cosmological constant is not included, because we focus only on black holes. In equation (\ref{eq31}) we confined all terms smaller than the logarithm into $\mathcal{O}(\Lambda)$, which can be expanded to make appear terms of the form $A^{-k}$ (with $k$ a positive integer). But according to \cite{Calmet_2021} we can interpret them as higher-loop terms in the effective action of the form (for the Schwarzschild case, otherwise there are more terms):
\begin{equation}\label{eq40}
    I^{(3)+}[g]=\int_\mathcal{M} d^4x \sqrt{-g}\left[ \frac{1}{128\pi^3} \left( c_6(\mu) R^{\mu \nu}{}_{\alpha \beta}R^{\alpha \beta}{}_{\rho \sigma}R^{\rho \sigma}{}_{\mu \nu}-\delta \, R^{\mu \nu}{}_{\alpha \beta} \ln \left( \frac{\square}{\mu^2} \right)R^{\alpha \beta}{}_{\rho \sigma}R^{\rho \sigma}{}_{\mu \nu} \right) +\mathcal{O}(R^4)\right].
\end{equation}
Using a non-zero cosmological constant, one can find \cite{pourhassan2022quantumgravitationalcorrectionsgeometry} that (\ref{eq39}) implies the following metric:
\begin{align}\label{eq59-60-61}
    ds^2 =& -f(r)dt^2+\frac{1}{g(r)}dr^2+r^2d\Omega^2
\\
    f(r) =& 1-\frac{r_s}{r}-\frac{\Lambda r^2}{3}-8\kappa \Lambda (4\alpha+\beta)\ln(r)
\\
    g(r) =& 1-\frac{r_s}{r}-\frac{\Lambda r^2}{3}-8\kappa \Lambda (4\alpha+\beta).
\end{align}
Upon inserting this metric into $I[g]$ (with a cosmological constant), the entropy is approximately of the form:
\begin{align}\label{eq62}
    S \stackrel{\Lambda \ll 1}{\simeq}& \frac{A}{4}+c_3(\mu)+\gamma \left( \ln\left( \frac{A\mu^2}{4\pi}\right)-2+2\gamma_\text{E} \right) + 128\kappa \Lambda (4\alpha+\beta)\left[ 6c_1(\mu)+c_2(\mu)-2c_3(\mu)\vphantom{\frac{0}{0}} \right.
\nonumber \\    
    & \left.+2(6\alpha+\beta-2\gamma)\left(\frac{1}{2}\ln\left( \frac{A\mu^2}{4\pi} \right)+2\gamma_\text{E}\right)\right],
\end{align}
because in \cite{pourhassan2022quantumgravitationalcorrectionsgeometry} this result is based on the approximation $r_h \simeq r_s$, where $r_h$ is the true radius of the black hole. The only way to make appear a $\Lambda$ term is to consider the $\mathcal{O}(\Lambda)$ in the entropy. It is possible to get rid of the $\Lambda \ln\left( \frac{A}{4}\right)$ term by defining either the triplet $(\alpha,\beta,\gamma)$ to suppress it, or the triplet $(c_1(1),c_2(1),c_3(1))$ to add the exact opposite contribution. By defining:
\begin{align}\label{eq63-64-65}
    c_1(\mu) =& c_1(1)-\alpha \ln\left( \frac{\mu^2}{M^2_\text{Pl}} \right)
\\
    c_2(\mu) =& c_2(1)-\beta \ln\left( \frac{\mu^2}{M^2_\text{Pl}} \right)
\\
    c_3(\mu) =& c_3(1)-\gamma \ln\left( \frac{\mu^2}{M^2_\text{Pl}} \right),
\end{align}
and considering the contribution of the cosmological constant in our entropy (\ref{eq31}), which is of the form $-\frac{4\Lambda}{\pi}+\mathcal{O}(\Lambda^2)$, we obtain the equation:
\begin{equation}\label{eq66}
    -\frac{4\Lambda}{\pi}=1024 \pi \Lambda (4\alpha+\beta)\left[ 6c_1(1)+c_2(1)-2c_3(1) +2(6\alpha+\beta-2\gamma)\left(\frac{1}{2}\ln\left( \frac{A}{4\pi} \right)+2\gamma_\text{E}\right)\right].
\end{equation}
The constants (\ref{eq63-64-65}) ensure that the entropy is generally renormalization group-invariant. We make the choice $6\alpha+\beta-2\gamma=0$. Thus, we are left with two degrees of freedom, for example, $(\alpha,c_1(1))$, we are free to choose to make the Wald entropy have the same form as our entropy. In general, higher loop corrections involve more coefficients than our entropy. This means our theory corresponds to a peculiar choice of coefficients. All this was for a free scalar field. But to include interactions, we need to take into account section \ref{ss3.2}. In this section, we derived the Liouville expectation value of a four-legged vertex with four associated propagators in position space. The result (\ref{eq27}) was a warm-up for the following section, where we consider a general propagator $\mathcal{G}$ as a series of propagators involving an increasing number of loops. We will not consider interactions between these general propagators $\mathcal{G}$, to simplify the calculations which are already quite heavy.

\subsection{Interacting scalar field}\label{ss4.2}

In this section, we directly use an interacting scalar field. A typical example of configuration of the momenta on the boundary is represented in the following illustration:
\begin{equation}\label{eq67}
    \includegraphics[scale=1.25]{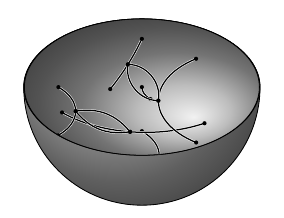}
\end{equation}
We will only consider the case where each line auto-interacts (that is to say, each line has its two ends on the boundary and has a given number of loops). The partition function is expressed by pulling out of the functional integral the sources, and then taking the expectation value $\langle \cdots \rangle_\text{K.}$ of the expression. Of course, we split the integration measure $\mathcal{D}h \leadsto \mathcal{D}(h_\text{cl.} \mathfrak{h})$ of the $H_3^+$-WZW model. Because evaluating the expectation value of the entire $\phi^4$ partition function is too hard to obtain, and because we have already given a vertex in a previous section, we focus on the propagator at the 1-loop order. The tadpole has the following structure:
\begin{align}\label{eq80}
    & \left\langle i\lambda 2i\beta_m \int d^3x \sqrt{\gamma} \,G(x_1,x)G(x,x)G(x_2,x)\right\rangle_\text{K.},\,\,\, \langle G(x,x) \rangle_\text{K.} \stackrel{!}{=} 1
\nonumber \\
    =& - \frac{\lambda}{2\beta_m} \int dr d^2 z\,r^2\left[\left\langle \varphi(z) \,\prod_{i=1}^2\left(e^{2\alpha_i \varphi(z_i)}e^{(\frac{Q}{2}-iP_i) \varphi(z)}\right)\prod_{k\notin \{i\}}e^{2\alpha_k \varphi(z_k)}\right\rangle_\text{L.}\right]^2
\nonumber \\
    \stackrel{\Lambda \rightarrow 0}{\rightarrow}& \,\sim -\frac{\lambda}{2\beta_m} \times \frac{-\ln(2)}{3} A \left( \frac{A}{\pi} \right)^{-2Q/b} 2^{-\frac{4m^2}{4\beta_m^2}\left(3r_s^2-3r_s \sum_i x_i^r+\frac{1}{2}((\sum_i x^r_i)^2+\sum_i {x^r_i}^2) \right)}\left( 2r_s-\sum_i x_i^r \right).
\end{align}
The corresponding diagram is of the following form:
\begin{equation}\label{eq81}
    \includegraphics[scale=0.35]{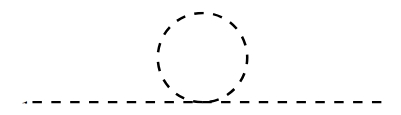}
\end{equation}
At two loops, the contributing diagrams are the sunset diagram and the double tadpoles. But from the correspondence between the Feynman diagram (\ref{eq81}) and its expectation value (\ref{eq80}), we can foresee some rules:
\begin{itemize}
    \item The coupling constant is changed, $i\lambda \leadsto \frac{-\lambda}{2\beta_m}$.
    \item Every expectation value comes with a factor $\left( \frac{A}{\pi} \right)^{-2Q/b}$ under our approximation.
    \item Each position integral gives a scaling factor $\propto A$.
\end{itemize}
The 2-loop order contains a diagram of the form:
\begin{equation}\label{eq82}
    \includegraphics[scale=0.35]{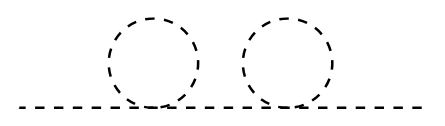}
\end{equation}
Evaluating its expectation value by brute force is tedious, so we rely on a trick of splitting the propagator's expectation value between the two vertices into two. First, we have:
\begin{align}\label{eq83}
    & \left\langle (i\lambda 2i\beta_m)^2\int d^3x\int d^3x' \sqrt{\gamma_x}\sqrt{\gamma_{x'}} \,G(x_1,x)G(x,x)G(x',x)G(x',x')G(x',x_2)\right\rangle_\text{K.}
\nonumber \\
    =& \frac{(-\lambda)^2}{2(2\beta_m)^2} \int dr\,r^2 \sum_l \int dr'\,{r'}^2 \sum_{l'} \left[\partial_\varepsilon \partial_{\varepsilon'} \left.\left\langle \,e^{(Q+\varepsilon-i\sum_i P_i)\varphi(z_l)}e^{(Q+\varepsilon'-i\sum_i P'_i)\varphi(z_{l'})}\prod_{k}e^{2\alpha_k \varphi(z_k)}\right\rangle_\text{L.}\right|_{\varepsilon,\varepsilon' =0}\right]^2.
\end{align}
The idea is then to split the propagator $G(x',x)\leadsto G(y,x)G(x',y)$. Indeed, we have its expectation value that is:
\begin{align}\label{eq84}
    \int_0^{r_s}dy^r\langle G(y,x)G(x',y) \rangle_\text{K.} =& \left[ \left\langle \left( e^{iP \varphi(z_y)}e^{-iP\varphi(z_x)} \right)\left( e^{iP' \varphi(z_{x'})}e^{-iP' \varphi(z_y)} \right)\prod_k e^{2\alpha_k \varphi(z_k)} \right\rangle_\text{L.} \right]^2
\nonumber \\
    \sim& \left( \frac{A}{\pi} \right)^{-2Q/b}\int_0^{r_s}dy^r\, 2^{-(P-P')^2-P^2-{P'}{}^2}
\nonumber \\
    \propto& \langle G(x',x) \rangle_\text{K.}.
\end{align}
Where the proportionality constant is $\Gamma^{-1} \equiv \frac{2\beta_m}{2m}\sqrt{\frac{\pi}{3\ln(2)}}$. In light of the rules enumerated above, it is clear that splitting our middle propagator leads to something of the form:
\begin{align}\label{eq85}
    & \int_0^{r_s}dy^r\left\langle (i\lambda 2i\beta_m)^2\int d^3x\int d^3x' \sqrt{\gamma_x}\sqrt{\gamma_{x'}} \,G(x_1,x)G(x,x)G(y,x)G(x',y)G(x',x')G(x',x_2)\right\rangle_\text{K.}
\nonumber \\
    \stackrel{\Lambda \rightarrow 0}{\sim}& \frac{(-\lambda)^2}{4\beta_m^2} \times \left(\frac{-\ln(2)}{3} A\right)^2 \left( \frac{A}{\pi} \right)^{-\frac{2Q}{b}} \Gamma\int_0^{r_s} dy^r \prod_{Y=X,X'} \left[2^{-\frac{4m^2}{4\beta_m^2}\left(3r_s^2-3r_s \sum_i Y_i^r+\frac{1}{2}((\sum_i Y^r_i)^2+\sum_i {Y^r_i}^2) \right)}\left( 2r_s-\sum_i Y_i^r \right)\right]
\nonumber \\
    \equiv& \frac{(-\lambda)^2}{4\beta_m^2} \times \left(\frac{-\ln(2)}{3} A\right)^2 \left( \frac{A}{\pi} \right)^{-\frac{2Q}{b}} \Gamma\int_0^{r_s} dy^r F^{(2)}_{r_s}[x_1^r,y^r]F^{(2)}_{r_s}[y^r,x_2^r].
\end{align}
It is then easy to generalize the one-loop approximation to the propagator (we write this generalization as $\mathcal{G}(x_1,x_2)$) because it is in the form of a series. Upon defining the zeroth order as just $\langle G(x_1,x_2)\rangle_\text{K.}$, we find:
\begin{equation}\label{eq86}
    \langle \mathcal{G}(x_1,x_2) \rangle_\text{K.} = \left( \frac{A}{\pi} \right)^{-\frac{2Q}{b}} \sum_{n=0}^\infty \left(\frac{-\lambda}{2\beta_m} \right)^n\left(\frac{-\ln(2)}{3} A\right)^n  \Gamma^{n}\int_0^{r_s} dy^r_1 \cdots dy^r_{n} F^{(2)}_{r_s}[x_1^r,y^r_1]F^{(2)}_{r_s}[y^r_1,y_2^r] \cdots F^{(2)}_{r_s}[y^r_{n},x_2^r].
\end{equation}
Finding a closed form for this sum seems impossible. However, we can take once again the advantage of the limit $\Lambda \rightarrow 0$, because it makes $\Gamma^2 F^{(2)}_{r_s}[r_s,x]$ very sharp, such that we can approximate it to be a derivative of a Dirac delta, centered at $r_s$. More precisely, the maximal variation of $F^{(2)}_{r_s}[r_s,x]$ occurs at $x=r_s-\frac{2\beta_m}{m\sqrt{8\ln(2)}}$. Thus, we make the approximation:
\begin{equation}\label{eq87}
    F^{(2)}_{r_s}[r_s,x] = e^{-\frac{4m^2 \ln(2)}{4\beta_m^2}(r_s-x)^2}(r_s-x) = -\frac{4\beta_m^2}{8m^2 \ln(2)} \frac{d}{dx}e^{-\frac{4m^2 \ln(2)}{4\beta_m^2}(r_s-x)^2}\stackrel{\Lambda \rightarrow 0}{\rightarrow} \frac{\sqrt{\pi}}{2} \left( \frac{2\beta_m}{2m \sqrt{\ln(2)}} \right)^3 \delta'(r_s-x).
\end{equation}
This approximation is a rough one because in fact, $0 < \Lambda \ll 1$, and we expect it to break at $n \sim \frac{1}{\sqrt{\Lambda}}$. But, assuming a coupling constant $0<\lambda <1$ should suppress these problematic contributions. With all this, we can approximate the $n$-th convolution power of $F^{(2)}_{r_s}[r_s,x]$ and find for the propagator at one loop:
\begin{equation}\label{eq88}
    \langle \mathcal{G}(r_s,x_2) \rangle_\text{K.}=\left( \frac{A}{\pi} \right)^{-\frac{2Q}{b}} \sum_{n=0}^\infty \left(-\frac{\lambda}{2\beta_m} \right)^n \left(\frac{-\ln(2)}{3} A\right)^n  \left( \frac{-\Gamma\sqrt{\pi}}{2} \right)^{n} \left( \frac{2\beta_m}{2m \sqrt{\ln(2)}} \right)^{3n} \frac{d^n}{dx_2^n}\left[e^{-\frac{4m^2 \ln(2)}{4\beta_m^2}(r_s-x_2)^2}\right].
\end{equation}
In what follows we write $\sigma = \frac{2\beta_m}{2m\sqrt{\ln(2)}}$, and so $\Gamma = \sigma^{-1}\sqrt{\frac{3}{\pi}}$. Thus the formula reduces to:
\begin{align}\label{eq89}
    \langle \mathcal{G}(r_s,x_2) \rangle_\text{K.} =& \left( \frac{A}{\pi} \right)^{-\frac{2Q}{b}} \sum_{n=0}^\infty \left(-\frac{\lambda}{2\beta_m} \right)^n \left(\frac{-\ln(2)}{3} A\right)^n  \left( \frac{-\sqrt{3}}{2 \sigma} \right)^{n} \sigma^{3n} \frac{d^n}{dx_2^n}\left[e^{-\frac{1}{\sigma^2}(r_s-x_2)^2}\right]
\nonumber \\
    =& \left( \frac{A}{\pi} \right)^{-\frac{2Q}{b}} \sum_{n=0}^\infty \chi^n \left(\frac{\sigma}{2\beta_m}\right)^{n} e^{-\frac{1}{\sigma^2}(r_s-x_2)^2} H_n\left( \left|\frac{r_s-x_2}{\sigma}\right| \right)
\nonumber \\
    =&\left( \frac{A}{\pi} \right)^{-\frac{2Q}{b}} \frac{\sqrt{\pi}}{2\varsigma\chi} e^{\frac{1-4|p| \sqrt{\ln(2)} \varsigma \chi}{4\varsigma^2\chi^2}}\mathrm{erfc}\left( \frac{1}{2 \varsigma \chi}-|p|\sqrt{\ln(2)}\right),\,\,\,\varsigma \equiv \frac{\sigma}{2\beta_m}=\left(2m\sqrt{\ln(2)}\right)^{-1}.
\end{align}
With $\chi \equiv \frac{\ln(2)}{2\sqrt{3}} \lambda A$. Furthermore, we have used the Hermite polynomials in the sum to obtain the middle expression. When $|\varsigma \chi| \rightarrow 0$, we find the free propagator again, indicating that the free theory is recovered when $\lambda=0$. In the previous subsection \ref{ss4.1}, the propagator had nice properties inherited from its exponential form, but one should note that this is not the case here. However, using the limit $|\varsigma \chi| \ll 1$, we can see that taking the product over the momenta is formally equivalent to replacing $|p| \leadsto |P| \equiv \sqrt{\sum_i p_i^2}$. Furthermore, we need to add a $A$-scaling factor to $P^2$, to be sure it depends on the horizon's area. Indeed, the product of propagators should depend on $A$ because this very product takes up to $\frac{A}{8}$ momenta. This trick is implemented by multiplying $|P|$ by $\sqrt{\frac{A}{8}}$ and taking the $\frac{8}{A}$-th root of the resulting propagator. Of course, in the limit $|\varsigma \chi| \rightarrow 0$ we obtain again a product of free propagators. We thus obtain the following entropy in the limit $A\rightarrow \infty$:
\begin{equation}\label{eq90}
    S^{1-\text{loop}} \stackrel{A \rightarrow \infty}{\sim} S^{0-\text{loop}}+ \frac{4}{A}\left[ 4+3\ln \left(\frac{3m^2}{\pi\ln(2)\lambda^2}\right)-6\ln\left(\frac{A}{4\pi}\right) \right]+\mathcal{O}(A^{-2}).
\end{equation}
Note that the coefficients $c_3$ and $\gamma$ are unchanged. The term in (\ref{eq90}) proportional to $A^{-1}$ can be rewritten as:
\begin{equation}\label{eq91}
    \frac{4}{A}\left[ 4+3\ln \left(\frac{3m^2}{\pi\ln(2)\lambda^2}\right)-6\ln\left(\frac{A}{4\pi}\right) \right] = c_6(\mu) \frac{1}{A}+\delta \frac{1}{A} \ln\left( \frac{A}{4\pi} \mu^2\right).
\end{equation}
We can see that $\delta = -24$. The entropy \textit{a la} Barvinsky-Vilkovisky is of the form:
\begin{align}\label{eq92}
    S_\text{BV} = \frac{A}{4}+c_3(\mu,\{p_k\}_k)+\gamma \left[ \ln\left( \frac{A\mu^2}{4\pi} \right)+2-2\gamma_\text{E}\right]+\frac{1}{A} \left[c_6(\mu)+\delta \ln\left( \frac{A\mu^2}{4\pi}\right) \right]+\mathcal{O}(A^{-2}).
\end{align} 
Of course, we have neglected the terms coming from the presence of a cosmological constant, as we make it tend to zero. Finally, we can give an expression for $c_6(1)$, which is:
\begin{equation}\label{eq93}
    c_6(1) \sim 16+12\ln \left(\frac{3m^2}{\pi\ln(2)\lambda^2}\right).
\end{equation}
Notice that in the limit of large $A$, (\ref{eq91}) is negative, which means the Bekenstein bound still holds. Remember that each propagator has one of its endpoints attached to the horizon and that we used the approximation of small bare coupling of the bulk scalar field theory. We can conclude this section by observing that at the lowest approximation to one loop diagrams (that is to say, each propagator $\mathcal{G}$ is a series, and none of its terms is linked to any other propagator), and assuming our trick of ``one of the end-points is on the boundary'' for each propagator, reproduces an effective term of the form $\mathcal{R}^3$ in the EFT action, where $\mathcal{R}$ is the curvature of 3+1D space-time. In the next part, we focus on the information content of the black hole and on the time evolution of its entropy. Indeed, until now we only studied the interior of the horizon, but its geometry can source far-away particles, leading to a decrease in the mass of the black hole and hence its entropy.

\section{Black hole evaporation and information}\label{s5}

\subsection{Modifications to Hawking radiations}\label{ss3.3}

The radiation, supposedly carrying the black hole interior information, is encoded onto the cosmological boundary in the same way information inside a black hole horizon is encoded onto it. Let us use a massive scalar field $\Phi$ with the equation of motion $\left( \Delta^\text{3D} -m^2-\frac{1}{4}R \right) \Phi=0$, as before. We are seeking a way to describe this field on the cosmological horizon. That is to say, we are searching for an expression of our field that is independent of the geometry of the bulk, and only on the geometry on the horizon. Indeed in the same way the expectation value of a propagator is expressed through the Liouville model on the boundary, we expect $\Phi$ to be expanded into modes carrying information on the boundary. It turns out that this is possible if one takes the operator $\left\langle \Delta^\text{3D} -m^2-\frac{1}{4}R \right\rangle_\text{K.}$ instead. Expressing the Klein-Gordon operator as the trace of the square of the Dirac operator in curved space, we obtain:
\begin{equation}\label{eq32.1}
    \Delta^\text{3D} -m^2-\frac{1}{4}R = \frac{1}{4} \text{tr}\left[ (i\slashed{D})(i\slashed{D})-m^2 \right].
\end{equation}
The Dirac operator is written as an exponential, to make the expectation value easier to deal with. 
\begin{equation}\label{eq32.2}
    i\slashed{D}=\sum_{i,d} \int dA_1 dA_2 \int d\chi d\omega\,\chi_d \omega^i \int d\xi d\overline{\xi}\,e^{\frac{i}{4\sqrt{\Lambda}}[A^+-A^-]^c_j (A_{1,c}A^j_2+\chi_c \omega^j)}e^{\overline{\xi} \left[ \gamma^d \left( \partial_i+\frac{1}{8}(A^++A^-)^a_i \epsilon^{abc}\sigma^{bc} \right) \right] \xi}.
\end{equation}
Where $A_1,A_2$ are bosonic, and the rest of the dummy variables are Grassmann-odd-valued. Our method allows us to take the expectation value over the two Chern-Simons theories, and from what we have done in \ref{ss3.1}, it is clear that the expectation value of the square will be the square of the expectation value. So we have:
\begin{align}\label{eq32.3}
    &\left\langle \Delta^\text{3D} -m^2-\frac{1}{4}R \right\rangle_\text{K.} 
\nonumber \\    
    \leadsto& -\frac{4 \Lambda}{\varphi(z_x)^2}\left[ \gamma^{(3)}\partial_r\pm i\varphi(z_x) \gamma^{(3)} \epsilon^{(3)ab}\sigma^{ab}\right]\left[ \gamma^{(3)}\partial_r\pm i\varphi(z_x) \gamma^{(3)} \epsilon^{(3)cd}\sigma^{cd}\right]-m^2
\nonumber \\    
    =& -\frac{4 \Lambda}{\varphi(z_x)^2}\left[\partial_r^2\pm i \varphi(z_x)(\gamma^{(3)}\gamma^{(3)} \epsilon^{(3)cd}\sigma^{cd}+\gamma^{(3)} \epsilon^{(3)ab} \sigma^{ab} \gamma^{(3)})\partial_r - \varphi(z_x)^2 \gamma^{(3)} \epsilon^{(3)ab}\sigma^{ab} \gamma^{(3)} \epsilon^{(3)cd}\sigma^{cd} \right]-m^2.
\end{align}
Thus, taking the trace of this matrix gives the expectation value of the Klein-Gordon operator with non-minimal coupling. Let $\Phi(x)\equiv \int \frac{dk}{2\pi}\left( \hat{a}{}^\dagger_k f^{}_k(z_x)+\hat{a}{}^{}_k f^\ast_k(z_x) \right)$ be the mode expansion of our field. Because here $k$ is a function of $x^r$, the equation of our modes $f_{k}(x)$ is:
\begin{align}\label{eq32.4}
    & \left(\partial_r^2+ \frac{1}{4}\varphi(z_x)^2 +\frac{m^2}{4\Lambda}\varphi(z_x)^2 \right)f_{k}(z_x) = 0 \Longrightarrow f_{k}(z_x) \stackrel{\Lambda \ll m^2}{\rightarrow} C_\pm\, e^{\pm i\frac{\Delta x^r m}{2\sqrt{\Lambda}}\varphi(z_x)}.
\end{align}
Where $\Delta x^r = x^r-y^r$, with $y^r$ independent of $x^r$. As before, we write $k=\frac{\Delta x^r m}{2\sqrt{\Lambda}}$. Now, we know the spectrum in the absence of sources (the sources are the poles in the geometry of the black hole horizon) is thermal \cite{Hawking1975}. So we expect the spectrum with sources to be a displaced thermal spectrum: a thermal coherent state. A coherent state with source $J$ is of the form:
\begin{equation}\label{eq32.5}
    | J \rangle \equiv e^{\int d^2x \sqrt{\sigma} J(x)\hat{\pi}(x)}|0\rangle,\,\,\, \hat{\pi}(x) \equiv \int \frac{d k}{2\pi} \left( \hat{a}{}^\dagger_k e^{-i k \varphi(z_x)}- \hat{a}_k e^{i k \varphi(z_x)} \right).
\end{equation}
We define the source to be $J(x) \equiv \frac{1}{V_\infty}\times A^{(3)}_z(z_x)|_{f^H=0}$, with $V_\infty$ the volume defined by the horizon for a faraway referential. The integral on $x$ is thus inversely proportional to the Schwarzschild radius $r_s$, so the larger the black hole, the lesser it sources particles. A standard calculation shows that the thermal expectation value of the number of excitations with momentum $q$ is: 
\begin{align}\label{eq32.6}
    \langle n_q(\varphi) \rangle_\text{th.} =& \int d^2x \sqrt{\sigma} [J(z_x) e^{i q \varphi(z_x)}]\int d^2y \sqrt{\sigma} [J(z_y) e^{-i q \varphi(z_y)}]+\frac{1}{e^{\beta \epsilon_q}-1}
\nonumber \\
    =& \frac{16}{V_\infty^2} \sum_{i=1}^N \sum_{j=1}^N \alpha_i \alpha_j e^{iq \varphi(z_i)}e^{-iq \varphi(z_j)}+\frac{1}{e^{\beta \epsilon_q}-1},\,\,\, N \equiv \frac{A}{4}.
\end{align}
If and only if a particle of momentum $q$ is inside the $f^H=0$ (closed) surface. Otherwise, there is only the second term, the Bose-Einstein distribution. We will use the approximation of the classical contribution of the partition function. That is to say, we take the regularized Liouville action \cite{Li_2020} with appropriate boundary conditions and leave the momenta unchanged by regularization, just as we have always done here. This essentially amounts to neglecting the correlation between the different $z_i$. Thus, we obtain:
\begin{align}\label{eq32.7}
    \left\langle \langle n_q(\varphi) \rangle_\text{th.} \right\rangle_\text{L.} =& \frac{16}{V_\infty^2}\frac{1}{Z[\alpha_k|z_k]}\sum_{i=1}^N \sum_{j=1}^N \alpha_i \alpha_j \left[\left\langle e^{2\alpha_q \varphi(z_i)}e^{2\alpha^\ast_q \varphi(z_j)} \prod_{k\neq i,j}^N e^{2\alpha_k \varphi(z_k)}\right\rangle_\text{L.} \right]^2+\frac{1}{e^{\beta \epsilon_q}-1}
\nonumber \\
    \sim& 36 \times \frac{(2\pi)^3}{\beta^4} 2^{-q^2}+\frac{1}{e^{\beta \epsilon_q}-1}.
\end{align}
Where we used the fact that the momenta cancel each other in each sum. Taking into account the condition ``Are there particles of momentum $k_i$ in the horizon'' (that is to say, whether there are scalar point particles inside of the boundary, or scalar propagators inserted in this region of space-time), we obtain the expectation value of the normal ordered Hamiltonian where $\overline{\sigma}$ is the radiation density constant (four times the Stefan-Boltzmann constant), and $T_\mathrm{BH} = \beta^{-1}$:
\begin{align}\label{eq32.8}
    \langle\langle:H:\rangle_\text{th.}\rangle_\text{L.} \sim& \int \frac{d^3k}{(2\pi)^3}\epsilon_k \left[ \vphantom{\frac{0}{0}} \right. \left(36\,\overline{\sigma}{}^{-1}(2\pi)^3\right)\left( \vphantom{\frac{0}{0}} \right. \underbrace{\overline{\sigma} T_\mathrm{BH}^4-\frac{m^2}{6}T_\mathrm{BH}^2}_{= \int \frac{d^3k}{(2\pi)^3} \frac{\epsilon_k}{e^{\beta \epsilon_k}-1}}+\frac{m^2}{6}T_\mathrm{BH}^2 \left. \vphantom{\frac{0}{0}} \right) \sum_i 2^{-k_i^2}\delta(k-k_i)+\frac{1}{e^{\beta \epsilon_k}-1} \left. \vphantom{\frac{0}{0}} \right]
\nonumber \\
    \stackrel{!}{=}& \int \frac{d^3k}{(2\pi)^3}\frac{\epsilon_k}{e^{\beta \epsilon_k}-1} \left[ 1 + \frac{540}{\pi^2 E_\mathrm{Pl}}\left( 1+\frac{m^2}{k^2} \right)\sum_{\{k_i\}} g_{k_i} \epsilon_{k_i} 2^{-k^2_i}\right].
\end{align}
Where $g_{k_i}$ is the degeneracy of the momentum $k_i$. Thus, we obtain a quantum-corrected number of particles. This indicates that information might be recovered in Hawking radiations because the spectrum associated with (\ref{eq32.8}) is a (corrected) thermal spectrum with pure emission lines. Note that for propagators (thus unconstrained in $k_i$), the quantum corrections are very small, while for scalar point particles (constrained by $0<\frac{k_i}{m}<1$) they are large.

\subsection{A word on the Page curve}\label{ss4.3}

\begin{figure}[ht]
    \centering
    \includegraphics[scale=0.75]{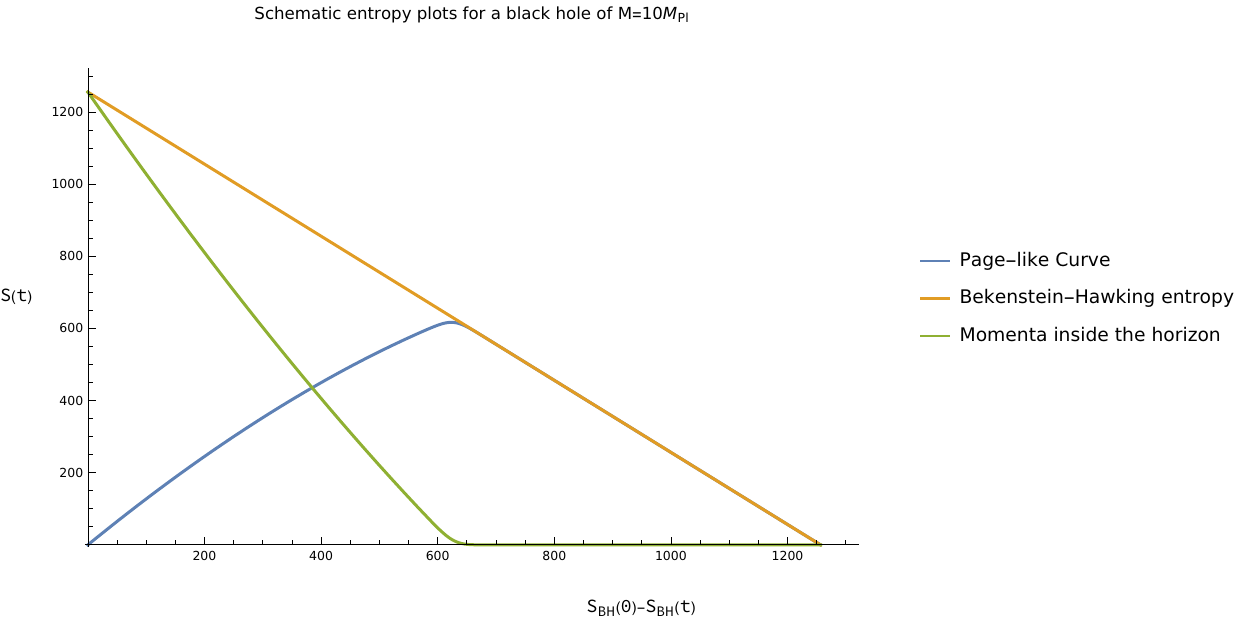}
    \caption{Schematic plot of our Page-like curve (in blueish color) versus the Bekenstein-Hawking entropy (in orange). The momentum content of the horizon is plotted in green and agrees with our assumptions.}
    \label{figure1}
\end{figure}

The main paradigm for finding the Page curve \cite{Page_2013}, is the island paradigm, where roughly speaking, a region inside the horizon is included in the non-classical part of the entropy. Then, the Page curve should be recovered using the Ryu-Takayanagi formula \cite{Ryu_2006}. In our case, the set-up is entirely different from that of Ryu and Takayanagi, and it seems that knowing the time evolution of $P^2 = \sum p^2$ can lead to a curve resembling closely the Page curve. Our argument is heuristic, and our approximations are based on subsection \ref{ss3.3}. In this subsection, we derived the corrections to the thermal spectrum, which are essentially pure emission rays on top of it. We will assume that in the entropy curve $S=f\left( S_\mathrm{BH} \right)$, the momenta inside of the horizon are radiated out at a constant rate throughout the evaporation, except at the Page time (the turning point of the Page curve), where eventually there is either a discontinuity or a sharp slope of the curve $\frac{d P^2}{dt}=f(t)$. The justification is the following: if the horizon is non-zero, but with no momenta, then $P^2=0$ constantly until the horizon evaporates completely. Furthermore, the maximum number of momenta is approximately given by ${P^{(0)}}^2>0$ in (\ref{eq32}). So there must be a sharp slope linking the behavior $\frac{d P^2}{dt} >0$ and $\frac{d P^2}{dt} =0$. Furthermore, our momenta come into pairs as our diagram (\ref{eqchord0}) shows. This means, that if one area bit $b_i$ containing $\ln(2)p_i^2$ evaporates, then $\ln(2)p_i^2$ will disappear from the other area bit $b'_i$. So we expect $\frac{d}{dt}\ln(2)P^2 \approx 2 \frac{dS_\mathrm{BH}}{dt}$ until $t=t_\mathrm{Page}$. \\
\indent With all this, we can plot an example of the time evolution of the momenta inside of the horizon as in Figure \ref{figure1}.
The green curve of Figure \ref{figure1} has the desired property: its derivative is almost everywhere constant. Furthermore, it fixes the Page time to be the time at which one-half of the Bekenstein-Hawking entropy remains. However, we should note that it is conceptually difficult to link $P^2(t)$ to the entanglement entropy of the radiated particles. One solution is to take the definition of the entanglement entropy as:
\begin{equation}\label{eq94}
    S_\mathrm{ent.} = (\beta \partial_\beta -1)(W_\mathrm{tot.}-W_\mathrm{grav.}).
\end{equation}
With $W_\mathrm{tot.} \sim W_\mathrm{grav.}+W$, where $W$ is the vacuum bubbles and the matter content, and $W_\mathrm{grav.}$ the gravitational action on the saddle point at which $\frac{\delta}{\delta g_{\mu \nu}}W_\mathrm{tot}=0$ (see \cite{Solodukhin_2011} for a review of the entanglement entropy of black holes). Our model is interesting because it is nearly in 3D (modulo a time factor $2\beta_m \ll 1$). This means that the vacuum bubbles of the matter will not involve terms higher than $\sim \mathcal{R}$ (where $\mathcal{R}$ is a curvature tensor). First, we need to be sure that the vacuum bubbles do not break our entropy calculation. We can use the zeta regularization to have $-\ln \det(D^2) = \zeta'_{D^2}(0)$, for $D^2=\Delta^\mathrm{3D}-m^2-\frac{1}{4}R$ and with:
\begin{align*}\label{eq95}
    \zeta_{D^2}(s) =& \frac{1}{\Gamma(s)} \int_{0^+}^\infty dt\,t^{s-1}\int d^3x \lim_{y\rightarrow x}\langle x | e^{tD^2} | y \rangle
\\
    =&\frac{1}{\Gamma(s)} \int_{0^+}^\infty dt\int d^3x \lim_{y\rightarrow x}\,t^{s-1} \frac{e^{-\frac{|x-y|}{2t}}}{(4\pi t)^\frac{3}{2}}\sum_{n\in \frac{1}{2} \mathbb{N}}  t^n a_{2n}(x,y)
\end{align*}
\begin{align}
    \leadsto& \lim_{|\epsilon| \rightarrow 0}\frac{1}{(4\pi)^\frac{3}{2}\Gamma(s)}\int d^3x \lim_{y\rightarrow x-\epsilon}\sum_{n\in \frac{1}{2} \mathbb{N}}  \frac{1}{\mu^{2n-3}}\left( \frac{\mu^2|x-y|^2}{2} \right)^{s+n-\frac{3}{2}}\Gamma\left(\frac{3}{2}-n-s\right)a_{2n}(x,y).
\end{align}
The HaMiDeW coefficients $a_{2n}(x) \equiv a_{2n}(x,x)$ are given in \cite{Vassilevich_2003}. The parameter $\mu$ is an energetic scale introduced so everything has the right unit. The derivative of the generalized zeta function on its argument is thus:
\begin{align}\label{eq96}
    \zeta'_{D^2}(s) \stackrel{|\epsilon| \rightarrow 0}{=}& -\int d^3x \sum_{n\in \frac{1}{2} \mathbb{N}} \frac{1}{\mu^{2n-3}} \left( \frac{|\epsilon|^2}{2\mu^{-2}} \right)^{s+n-\frac{3}{2}}\frac{\Gamma(\frac{3}{2}-n-s)}{(4\pi)^\frac{3}{2}\Gamma(s)}\left[ \psi\left(\frac{3}{2}-n-s\right)+\psi(s) -\ln\left( \frac{|\epsilon|^2}{2\mu^{-2}} \right) \right]a_{2n}(x).
\end{align}
Where $\psi$ is the logarithm derivative of the Euler $\Gamma$ function. We have boundaries and this justifies the sum on half-integer values. Thus, upon choosing the energetic scale $\mu$ such that the logarithmic term vanishes, and reintroducing the integration measures, we obtain ($K$ is the extrinsic curvature on $\partial \Sigma$):
\begin{equation}\label{eq97}
    \zeta'_{D^2}(0) = \frac{1}{(4\pi)^\frac{3}{2}}\int_{\Sigma} d^3x \sqrt{\gamma} \left[ \frac{1-m^2|\epsilon|^2}{|\epsilon|^3}-\frac{1}{12|\epsilon|}R \right]+\frac{1}{(4\pi)^\frac{3}{2}}\int_{\partial \Sigma} d^2x \sqrt{\sigma} \left[\frac{1}{4|\epsilon|^2}\sqrt{4\pi}+\frac{1}{3|\epsilon|}K \right].
\end{equation}
These terms can entirely be absorbed into the definition of $W_\mathrm{grav.}$, leading to renormalized constants ($G$, $\Lambda$, $\Lambda_\partial$ and $G_\partial$). Thus, only a term of the form $-\ln\left(\langle \prod_{k=1}^\mathcal{N} G(x_k,y_k) \rangle_{\mathrm{CS}^\pm}\right)$ remains. By our point particle/propagator duality, this term is equivalent to $\mathcal{N}$ point particles inside the black hole horizon. This means that, apart from a renormalization of the gravitational parameters, the entanglement entropy is:
\begin{equation}\label{eq98}
    S_\mathrm{ent.} = (1-\beta \partial_\beta)\ln\left(\left\langle \prod_{k=1}^\mathcal{N} G(x_k,y_k) \right\rangle_{\mathrm{CS}^\pm}\right) \simeq 8-4\ln \left( \frac{A}{\pi} \right)-\ln(2)P^2.
\end{equation}
With, as before, $P^2\equiv \sum p^2$. Thus, in this approach, we can identify the logarithmic term and the constant $8$ as coming from the entanglement side of the entropy, not from the gravitational part. However, in the Effective Field Theory approach, it comes from the gravitational part. This is just a change of paradigm. Note that the condition $W_\mathrm{grav.} = S_\mathrm{grav.}|_\text{saddle point}$ is consistent with what we have done because we split the integration measure of the $H_3^+$-WZW model into $\mathcal{D}h \leadsto \mathcal{D}(h_\mathrm{cl.} \mathfrak{h})$, as we have always done in this paper. This entropy is negative, and although it is interpreted as a quantum-field theoretic entanglement entropy, it remains a thermodynamical entropy. But it does not matter because this entropy is unphysical if taken alone. Indeed, it precludes a black hole of area $A$, so the total entropy $S_\mathrm{tot.}\equiv S_\mathrm{grav.}+S_\mathrm{ent.}$ is the only physical entropy. This implies that $S_\mathrm{grav.}$ is not physical either, and this explains why there is an apparent paradox of information: One cannot isolate the entanglement entropy from the gravitational entropy in a full theory of quantum gravity with matter. This is best illustrated by (\ref{eq30.2}) and (\ref{eq31}).\\
\indent The $8-4\ln \left( \frac{A}{\pi} \right)$ fully comes from conformal fluctuations on the boundary, while $-\ln(2)P^2$ is precisely the negative interior entropy associated with the exterior radiation. Indeed, let us take a simple model of free scalar field propagators saturating the horizon at instant $t=0$. We denote by $B$ the interior of the boundary/horizon (excluding it), and $\overline{B }$ its complement (the exterior and the horizon). Then, interpreting the end-points of a propagator as either ``ingoing'' or ``outgoing'' (which we write as $|\uparrow \rangle$ and $|\downarrow \rangle$ respectively) from the horizon, we obtain that the state of the scalar field on the boundary must be:
\begin{equation}\label{eq99}
    |\Psi \rangle = \bigotimes_{n=1}^\mathcal{N} \left[\frac{1}{\sqrt{2}}\left( | \uparrow \rangle_{B,n} \otimes | \downarrow \rangle_{\overline{B},n} + |\downarrow \rangle_{B,n} \otimes |\uparrow \rangle_{\overline{B},n} \right) \right].
\end{equation}
With the condition $\mathcal{N}(0)\equiv \mathcal{N}= \frac{A(0)}{4}$. This ought to be our state because in the Euclidean signature, there is no way to distinguish between $G(x,y)$ and $G(y,x)$ (while in Lorentzian we have a time-ordering distinguishing between $x^0 > y^0$ and $x^0< y^0$.) Thus, the entanglement entropy inside the horizon is:
\begin{align}\label{eq100}
    S_B =& -\mathrm{Tr}_{B}[\rho_B \ln(\rho_B)] = -2\times \frac{1}{2}\sum_{n=1}^{\mathcal{N}} \ln\left( \frac{1}{2} \right) = \mathcal{N} \ln(2).
\end{align}
So we identify $\mathcal{N}=P^2$, which means that $\ln(2)\mathcal{N}(t) \propto \max\left( 2S_\mathrm{BH}(t)-S_\mathrm{BH}(0),0 \right)$. The maximal entropy $\ln(2)$ of the scalar particles coinciding with $\ln(2\ell_\mathrm{Pl})$ may find its origin in the fact that our theory appears purely gravitational, even with a scalar field, as illustrated by (\ref{eq8}).

\section{Discussion and conclusion}\label{s6}

\indent In part \ref{s3}, we saw that one can derive the expectation value of a scalar propagator and find the same result as for a point source propagating. However, it is yet to be understood why (other than mathematically) this is the case because conceptually, a quantum-field theoretic propagator has nothing to do with point particles. It also remains to see if this duality persists for particles with spin. One possible explanation is that for the gravitational field, a propagator $G(x,y)$ with mass $m$ is the same as a particle of mass $m$ traveling along a geodesic linking $x$ and $y$. A possible implication is that maybe when we consider an initial field $\Phi$ configuration $\Phi(f)$ and a final field configuration $\Phi(g)$ (for $f$ and $g$ two test functions since $\Phi$ is technically a distribution), then the expectation value over all the geometries of space-time of $\langle 0 | \Phi(f) \Phi(g) |0 \rangle$ gives a weighted double sum over all pairs $(x,y)$ of a point particle contribution. The dominant contribution to this weighted double sum is thus in the union of the overlap of $f$ and $g$ (the weights), and the set of pairs $(x,y)$ such that $y^r \in [x^r-2\beta_m, x^r+2\beta_m]$. Assuming our test functions $f$ and $g$ are localized at $y$ and $x$ respectively, this means that in two different referentials --- the centers of genus-0 boundaries --- the dominant contribution to the propagator can be reflected by two different paths. Of course, this latter point is true only for the cosmological horizon because it is observer-dependant, while for black holes there is only one possible pair $(x,y)$.\\
\indent On the other side, by adding quartic interactions to the scalar field, we have found the expectation value of a truncated Feynman diagram inside the horizon of a black hole. This expectation value tells us that interacting theories have an interaction strength increasingly greater than the unity when we are close to the horizon (but still inside of it), while near the singularity, the field is nearly free. The same phenomenon rules the self-energy of a propagator. When calculating the geometric series of propagators with an increasing number of distinct loops, we obtain a new propagator in the limit where one of its ends lies on the horizon, meaning that the series converges.\\
\indent Calculating the true entropy is near impossible because of the enormous amount of vertex operator insertions in the Liouville expectation value. We used a crude approximation, based on the the classical approximation with renormalized action. In the process, and because the momentum insertions do not belong to the action, we did not include any counter-terms for them. We find an entropy that agrees with the Effective Field Theory approach to Quantum Gravity \cite{Donoghue_1994}: the Bekenstein-Hawking entropy is corrected by a logarithmic term plus a series of terms involving inverse powers of the area of the horizon. We have deduced that in the regime of low energetic scale, the model proposed in \cite{Roux:2024bna} coincides with a special case of Quantum Gravity treated as an effective field theory (EFT). In the absence of matter, the series of inverse powers of $A$ can be brought to a closed expression, so that the apparent poles $A=0$ are cancelled. However, we generally cannot expect such nice behavior to occur in the presence of matter, at least not in the form of a geometric series. The key difference between our low energy effective action and the one of EFT Quantum Gravity is that our coefficients are fixed. It is interesting to observe that according to \cite{ZHANG200914}, the entropy corrections are logarithmic and in the inverse power of the area of the horizon, just as our result (\ref{eq90}). The fact that, at least at order three in curvature, the corrections to the entropy are of negative signs indicates a non-violation of the Bekenstein bound.\\
\indent Finally, we found a quantum correction to the Hawking radiations. This correction is of the form of pure emission lines on top of the thermal spectrum. Furthermore, we tried to obtain a Page-like curve by assuming the horizon is initially saturated with particles. We then gave a heuristic derivation of how the total sum of the momenta squared should evolve with time. In this model, what we call ``entanglement entropy'' in QFT is translated into a term of the form $\sim 8-4\ln\left( \frac{A}{\pi}\right)+\ln(2^{-\sum p^2})$. If this is a valid approach, this would explain why we have easily found a Page-like curve, which would truly be the Page curve in this case. If this is true, this may be an indication that there exists a simpler method to obtain the Page curve than the current one involving the replica trick and Euclidean wormholes.\\
\indent We conclude by saying that not only does it seem possible to include interacting matter in the model presented in \cite{Roux:2024bna}, but also to find corrections to the emission spectrum of Hawking radiations and the numerical values of the different coefficients appearing in the Wald entropy of the EFT Quantum Gravity. Thus, the model of \cite{Roux:2024bna} is predictive.

\section*{Acknowledgements}

The author is grateful to the anonymous referees for their valuable comments on the readability of the initial version of the manuscript.

\divider
\section{References}
\nocite{*}
\bibliographystyle{iopart-num}
\bibliography{Bibliography}

\providecommand{\newblock}{}
\begin{thebibliography}{10}
\expandafter\ifx\csname url\endcsname\relax
  \def\url#1{{\tt #1}}\fi
\expandafter\ifx\csname urlprefix\endcsname\relax\def\urlprefix{URL }\fi
\providecommand{\eprint}[2][]{\url{#2}}

\bibitem{Roux:2024bna}
Roux J~B 2024 {\em Class. Quant. Grav.\/} {\bf 41} 147001

\bibitem{Alexandrov_1998}
Alexandrov S~Y and Vassilevich D~V 1998 {\em Physical Review D\/} {\bf 58} ISSN
  1089-4918 \urlprefix\url{http://dx.doi.org/10.1103/PhysRevD.58.124029}

\bibitem{Yildirim_2015}
Yildirim T 2015 {\em International Journal of Modern Physics A\/} {\bf 30}
  1550034 ISSN 1793-656X part 2
  \urlprefix\url{http://dx.doi.org/10.1142/S0217751X15500347}

\bibitem{Li_2020}
Li S, Toumbas N and Troost J 2020 {\em Nuclear Physics B\/} {\bf 952} 114913
  ISSN 0550-3213 part 4
  \urlprefix\url{http://dx.doi.org/10.1016/j.nuclphysb.2019.114913}

\bibitem{chatterjee2024liouvilletheoryintroductionrigorous}
Chatterjee S and Witten E 2024 Liouville theory: An introduction to rigorous
  approaches (\textit{Preprint} \eprint{2404.02001})
  \urlprefix\url{https://arxiv.org/abs/2404.02001}

\bibitem{gawedzki1999conformalfieldtheorycase}
Gawedzki K 1999 Conformal field theory: a case study (\textit{Preprint}
  \eprint{hep-th/9904145}) \urlprefix\url{https://arxiv.org/abs/hep-th/9904145}

\bibitem{corradini2021spinning}
Corradini O, Schubert C, Edwards J~P and Ahmadiniaz N 2021 Spinning particles
  in quantum mechanics and quantum field theory (\textit{Preprint}
  \eprint{1512.08694})

\bibitem{Zamolodchikov_1996}
Zamolodchikov A and Zamolodchikov A 1996 {\em Nuclear Physics B\/} {\bf 477}
  577–605 ISSN 0550-3213
  \urlprefix\url{http://dx.doi.org/10.1016/0550-3213(96)00351-3}

\bibitem{Giribet_2022}
Giribet G and Leoni M 2022 {\em Journal of High Energy Physics\/} {\bf 2022}
  ISSN 1029-8479 \urlprefix\url{http://dx.doi.org/10.1007/JHEP09(2022)126}

\bibitem{Donoghue_1994}
Donoghue J~F 1994 {\em Physical Review D\/} {\bf 50} 3874–3888 ISSN 0556-2821
  \urlprefix\url{http://dx.doi.org/10.1103/PhysRevD.50.3874}

\bibitem{Calmet_2021}
Calmet X and Kuipers F 2021 {\em Physical Review D\/} {\bf 104} ISSN 2470-0029
  \urlprefix\url{http://dx.doi.org/10.1103/PhysRevD.104.066012}

\bibitem{pourhassan2022quantumgravitationalcorrectionsgeometry}
Pourhassan B and Delgado R~C 2022 Quantum gravitational corrections to the
  geometry of charged ads black holes (\textit{Preprint} \eprint{2205.00238})
  \urlprefix\url{https://arxiv.org/abs/2205.00238}

\bibitem{Hawking1975}
Hawking S~W 1975 {\em Communications in Mathematical Physics\/} {\bf 43}
  199--220 ISSN 1432-0916 \urlprefix\url{https://doi.org/10.1007/BF02345020}

\bibitem{Page_2013}
Page D~N 2013 {\em Journal of Cosmology and Astroparticle Physics\/} {\bf 2013}
  028–028 ISSN 1475-7516
  \urlprefix\url{http://dx.doi.org/10.1088/1475-7516/2013/09/028}

\bibitem{Ryu_2006}
Ryu S and Takayanagi T 2006 {\em Physical Review Letters\/} {\bf 96} ISSN
  1079-7114 \urlprefix\url{http://dx.doi.org/10.1103/PhysRevLett.96.181602}

\bibitem{Solodukhin_2011}
Solodukhin S~N 2011 {\em Living Reviews in Relativity\/} {\bf 14} ISSN
  1433-8351 \urlprefix\url{http://dx.doi.org/10.12942/lrr-2011-8}

\bibitem{Vassilevich_2003}
Vassilevich D 2003 {\em Physics Reports\/} {\bf 388} 279–360 ISSN 0370-1573
  \urlprefix\url{http://dx.doi.org/10.1016/j.physrep.2003.09.002}

\bibitem{ZHANG200914}
Zhang J 2009 {\em Physics Letters B\/} {\bf 675} 14--17 ISSN 0370-2693
  \urlprefix\url{https://www.sciencedirect.com/science/article/pii/S0370269309003943}

\bibitem{Bhattacharya:2017wlw}
Bhattacharya S 2017 {\em Adv. High Energy Phys.\/} {\bf 2017} 2165731

\end{thebibliography}

\appendix
\section{}
\setcounter{section}{1}\label{appendix}

We would like to see if performing the same expectation value for a vector boson propagator is still possible. We chose a vector field directly because it is harder to introduce than a spinor field in practice. Indeed, the vector boson propagator can be expressed as an $N=2$ supersymmetric worldline formalism, while an $N=1$ supersymmetric theory describes a spinor propagator. First, let us show this statement. The BRST Hamiltonian action of the $N=2$ supersymmetric worldline formalism is given in \cite{Bhattacharya:2017wlw} to be:
\begin{equation}\label{eq20.1}
    S= \int_0^1 ds \left[ p_i \dot{x}^i +i\overline{\psi}_a \dot{\psi}{}^a+\dot{\mathcal{C}}{}^A \mathcal{P}_A-2TH -\vartheta (\overline{\psi}{}^a \psi_a+N_{\overline{\mathcal{C}}{}'}-N_{\mathcal{C}'})+s\vartheta\right].
\end{equation}
Where $H = \frac{1}{2}\pi_i \pi^i-\frac{1}{2}R_{abcd}\overline{\psi}{}^a \psi^b \overline{\psi}{}^c \psi^d$, with $\pi_i = p_i-i\omega_{iab}\overline{\psi}{}^a \psi^b$. The quantities $\mathcal{C}$ and $\mathcal{P}$ are the BRST ghosts and their conjugated momenta. The parameter $\vartheta$ is a constant $\text{U}(1)$ gauge field, and $s=\frac{D}{2}-p-1=-\frac{1}{2}$ because we are in $D=3$ dimensions, and we describe a $p=1$ form (the photon field). The propagator is thus given by the expression:
\begin{equation}\label{eq20.2}
    \int_0^\infty dT \int_0^{2\pi}\frac{d\vartheta}{2\pi} \int \mathcal{D} \mathcal{C} \int \mathcal{D} \mathcal{P} \int \mathcal{D}p \int_{x(0)=x}^{x(1)=y}\mathcal{D}x \sqrt{\det(\gamma)}\int_{\psi(0)=\psi_1}^{\overline{\psi}(1)=\overline{\psi}_2} \mathcal{D} \psi \mathcal{D} \overline{\psi} e^{-S_E+\overline{\psi}_a \psi^a(1)+\mathcal{C}_A \mathcal{P}^A(1)}.
\end{equation}
Where boundary terms have been introduced to be able to do the functional integrals correctly. From this same reference \cite{Bhattacharya:2017wlw}, we can integrate the BRST ghosts and their conjugated momenta to have:
\begin{align}\label{eq20.3}
    (\ref{eq20.2}) =& \int_0^\infty dT \int_0^{2\pi}\frac{d\vartheta}{2\pi}e^{e^{-i\vartheta} \overline{\psi}_2 \cdot \psi_1-i(s-2)\vartheta} \langle y | e^{-2TH} | x \rangle = \int_0^\infty dT \langle y, \psi_2 |e^{-2TH}| x, \psi_1 \rangle.
\end{align}
Integrating on $\vartheta$ gives an expectation value on the states $\langle y, \psi_2 |$ and $| x, \psi_1 \rangle$. To perform this expectation value, we expand in series the exponential of the Hamiltonian:
\begin{equation}\label{eq20.4}
    \langle y, \psi_2 |e^{-2TH}| x, \psi_1 \rangle = \langle y,0 |\psi_\mu \left[ 1-T\left( \pi_i \pi^i -R_{abcd}\overline{\psi}{}^a \psi^b \overline{\psi}{}^c \psi^d\right)+\cdots \right]\overline{\psi}_\nu| x,0 \rangle.
\end{equation}
But, we have the anti-commutation relations of the $0|2$ dimensional variables $\{\psi_\mu , \overline{\psi}_\nu\} = g_{\mu \nu} \Rightarrow \psi_\mu \overline{\psi}_\nu=g_{\mu \nu}-\overline{\psi}_\nu \psi_\mu$. Moreover, $\psi$ annihilates $|0\rangle$. So we have in fact:
\begin{align}\label{eq20.5}
    \langle y, \psi_2 |\frac{1}{2H}| x, \psi_1 \rangle = \langle y,0 | \frac{1}{\gamma_{ij} \nabla_k \nabla^k-R^\text{3D}_{ij}} | x,0 \rangle.
\end{align}
Which is indeed the propagator of a photon in 3D curved space. Now, we can proceed further from (\ref{eq20.2}). When integrating on $p$ in (\ref{eq20.2}), we obtain the following action in Euclidean signature:
\begin{equation}\label{eq20.6}
    S'_E= \int_0^1 ds \left[ \frac{1}{4T}\gamma_{ij} \dot{x}{}^i \dot{x}{}^j -\dot{\mathcal{C}}{}^A \mathcal{P}_A-TR_{abcd}\overline{\psi}{}^a \psi^b\overline{\psi}{}^c \psi^d+ \overline{\psi}_a (\dot{\psi}{}^a+ \dot{x}{}^i \omega_{i}{}^{ab} \psi_b) -i\vartheta (\overline{\psi}{}^a \psi_a+N_{\overline{\mathcal{C}}{}'}-N_{\mathcal{C}'})+si\vartheta\right].
\end{equation}
We are specifically interested in the part $\int_0^1 ds \left[ \frac{1}{4T}\gamma_{ij} \dot{x}{}^i \dot{x}{}^j +\dot{x}{}^i \omega_{i}{}^{ab}\overline{\psi}_a \psi_b - TR_{abcd}\overline{\psi}{}^a \psi^b\overline{\psi}{}^c \psi^d \right]$, because this is the relevant one when doing the Chern-Simons expectation value. The first term ($\gamma_{ij} \dot{x}{}^i \dot{x}{}^j$) has already been treated in the previous section. We must express the spin connection $\omega$ and the Riemann tensor $R_{abcd}$ as a function of the gauge fields $A^+$ and $A^-$, as before. Note that no counterterms are arising from the Lee-Yang ghosts method for this model, so the Riemann tensor stays here uncompensated. We have:
\begin{align}\label{eq20.7}
    & \omega_i{}^{ab} = \frac{1}{4}\epsilon_{c}{}^{ab}(A^{+,c}_i+A^{-,c}_i),
\nonumber \\
    & R_{ijcd}\overline{\psi}{}^i \psi^j\overline{\psi}{}^c \psi^d = \left( \partial_i \omega_{jcd}-\partial_j \omega_{icd}+\omega_{ic}{}^f \omega_{jfd}-\omega_{jc}{}^f \omega_{ifd}\right)\overline{\psi}{}^i \psi^j \overline{\psi}{}^c \psi^d.
\end{align}
Dealing with the Riemann tensor is hard if one keeps it on this side of the propagator (inside of the functional integrals). To make the calculation easier, we pull out of the functional integrals the Riemann tensor, by introducing sources and an operator $\mathcal{O}$:
\begin{equation}\label{eq20.8}
    \mathcal{O}\equiv \left.\exp\left(T \int_0^1 ds R_{ijcd}\left( \frac{\delta}{\delta J^k_f}, \frac{\delta}{\delta K^k_f} \right) \frac{\delta}{\delta \eta_i}\frac{\delta}{\delta \overline{\eta}_j} \frac{\delta}{\delta \eta_c}\frac{\delta}{\delta \overline{\eta}_d}\right) \bullet \right|_{J,K,\eta,\overline{\eta}=0}.
\end{equation}
Note the difference in indices between this Riemann tensor and the one appearing in (\ref{eq20.6}). To change the flat indices into curved ones, we must use the tetrads, so we have to introduce sources for the tetrads too. The complete source term to be introduced in the action is $J^i_a e_i^a+K^i_a \omega_i^a+\eta_a \overline{\psi}{}^a+\overline{\eta}_a \psi^a$. This action can be split into two parts, one involving the tetrads and the spin connections, and another one involving the remaining quantities:
\begin{align}\label{eq20.9-10}
    S_1 =& \int_0^1 ds \left[ \frac{1}{4T}\gamma_{ij} \dot{x}{}^i \dot{x}{}^j + \frac{1}{4} \dot{x}{}^i \epsilon_{c}{}^{ab}(A^{+,c}_i+A^{-,c}_i) \overline{\psi}_a \psi_b +\frac{1}{2\sqrt{\Lambda}}J_a^i\left[ A^{+,a}_i-A^{-,a}_i \right] + \frac{1}{2}K^i_a \left[ A^{+,a}_i+A^{-,a}_i \right]\right]
\\
    S_2 =& \int_0^1 ds \left[ -\dot{\mathcal{C}}{}^A \mathcal{P}_A + \overline{\psi}_a \dot{\psi}{}^a +\eta_a \overline{\psi}{}^a+\overline{\eta}_a \psi^a-i\vartheta (\overline{\psi}{}^a \psi_a+N_{\overline{\mathcal{C}}{}'}-N_{\mathcal{C}'})+si\vartheta\right].
\end{align}
Analogously to the equation (\ref{eq16}), we introduce a wave-functional $\Psi[A_z|_\partial]$ arising from the usual method for finding the Chern-Simons expectation value from WZW ones. This wave-functional obeys the equation:
\begin{equation}\label{eq20.11}
    \left( \frac{k}{4\pi} F^\pm_{z\overline{z}}-[\cdots] - \int_0^1 ds \left(\pm \frac{1}{8}\dot{x}{}^r \epsilon_c{}^{ab} \sigma^{\text{T},c} \overline{\psi}_a \psi_b + \frac{1}{4\sqrt{\Lambda}} J^r_c \sigma^{\text{T},c} \pm \frac{1}{4} K^r_c \sigma^{\text{T},c}\right)\delta^{(3)}(x(s)-x)\right) \Psi[A_z|_\partial]=0.
\end{equation}
Where $[\cdots]$ is the same contribution involved in (\ref{eq16}). Thus, we deduce the following expression for the propagator:
\begin{equation}\label{eq20.12}
    \int_0^\infty dT \int_0^{2\pi}d\vartheta \int \mathcal{D} \mathcal{C} \int \mathcal{D} \mathcal{P} \int_{x^r(0)=x^r}^{x^r(1)=y^r}\mathcal{D}x \sqrt{\det[\varphi(z_1)-\varphi(z_0)]^2}\int_{\psi(0)=\psi_1}^{\overline{\psi}(1)=\overline{\psi}_2} \mathcal{D} \psi \mathcal{D} \overline{\psi} e^{-S^{\pm}-S'-\overline{\psi}_a \psi^a(1)-\mathcal{C}_A \mathcal{P}^A(1)}.
\end{equation}
In this expression, the action $S^\pm$ contains all the contributions from the gauge fields $A^\pm$ while $S'$ contains the remaining terms. Specifically, we have:
\begin{equation}\label{eq20.13}
    S^\pm_1 = \int_0^1 ds \left[ \frac{1}{16\Lambda T} [\varphi(z_1)-\varphi(z_0)]^2 \dot{x}{}^r \dot{x}{}^r \pm \frac{1}{4} \dot{x}{}^r \epsilon_{3}{}^{ab}[\varphi(z_1)-\varphi(z_0)] \overline{\psi}_a \psi_b + \frac{1}{2}\left[\frac{1}{\sqrt{\Lambda}}J^r_3 \pm K^r_3 \right][\varphi(z_1)-\varphi(z_0)]\right].
\end{equation}
Upon introducing the operator $\mathcal{O}' \equiv \left.\exp\left(\mp \frac{1}{4} \int_0^1 ds \dot{x}{}^r \epsilon_{3}{}^{ab}[\varphi(z_1)-\varphi(z_0)] \frac{\delta}{\delta \eta^a} \frac{\delta}{\delta \overline{\eta}{}^b} \right) \bullet \right|_{\eta, \overline{\eta}=0}$ to be able to deal with the middle term, we obtain a new expression for the propagator of a photon:
\begin{align}\label{eq20.14}
    & \int_0^\infty dT \int_0^{2\pi}\frac{d\vartheta}{2\pi}e^{e^{-i\vartheta} \overline{\psi}_2 \cdot \psi_1-i(s-2)\vartheta} \mathcal{O}\mathcal{O}'e^{i\int_0^1 ds e^{-i\vartheta} \overline{\psi}{}^a_{2}\eta_a+i\int_0^1 ds \psi^a_{1}\overline{\eta}_a+\int_0^1 ds \int_0^1 ds' \overline{\eta}_a(s) G^{ab}(s-s')\eta_b(s')}  
\nonumber \\    
    & \times \int_{x^r(0)=x^r}^{x^r(1)=y^r}\mathcal{D}x \sqrt{\det [\varphi(z_1)-\varphi(z_0)]^2} e^{-\int_0^1 ds \left(\frac{1}{16\Lambda T} [\varphi(z_1)-\varphi(z_0)]^2 (\dot{x}{}^r)^2+\frac{1}{2}\left[\frac{1}{\sqrt{\lambda}}J^r_3 \pm K^r_3 \right][\varphi(z_1)-\varphi(z_0)] \right)}.
\end{align}
With $G^{ab}(s-s') = \delta^{ab}\Theta(s-s')$. We perform a series expansion of the operator $\mathcal{O}'$. Its contribution is:
\begin{align}\label{eq20.15}
    & \left.\left(1\mp \frac{1}{4} \int_0^1 ds \dot{x}{}^r \epsilon_{3}{}^{ab}[\varphi(z_1)-\varphi(z_0)] \frac{\delta}{\delta \eta^a} \frac{\delta}{\delta \overline{\eta}{}^b} +\cdots \right)e^{i\int_0^1 ds e^{-i\vartheta} \overline{\psi}{}^a_{2}\eta_a+i\int_0^1 ds \psi^a_{1}\overline{\eta}_a+\int_0^1 ds \int_0^1 ds' \overline{\eta}_a(s) G^{ab}(s-s')\eta_b(s')} \right|_{\eta,\overline{\eta}=0}
\nonumber \\
    &= \left(1\mp \frac{1}{4} \int_0^1 ds \dot{x}{}^r \epsilon_{3}{}^{ab}[\varphi(z_1)-\varphi(z_0)] e^{-i\vartheta}\overline{\psi}_{2,a} \psi_{1,b} +\cdots \right).
\end{align}
As we can see, the contribution purely due to the spin connection does not vanish. Next, the action of the operator $\mathcal{O}$ is, after doing a series expansion:
\begin{align}\label{eq20.16}
    &\left. \mathcal{O} e^{i\int_0^1 ds e^{-i\vartheta} \overline{\psi}{}^a_{2}\eta_a+i\int_0^1 ds \psi^a_{1}\overline{\eta}_a+\overline{\eta}_p \ast G^{pq} \ast \eta_q}e^{\int_0^1 ds \frac{1}{2}\left[\frac{-1}{\sqrt{\Lambda}}J^r_3 \mp K^r_3 \right]\Delta \varphi} \right|_{J,K,\eta,\overline{\eta}=0}
\nonumber \\
    =& \left.\left(1+T \int_0^1 ds \frac{\delta}{\delta J_a^i} \frac{\delta}{\delta J_b^j} R_{ijcd}\left( \frac{\delta}{\delta K^k_f} \right) \left[ e^{-2i\vartheta} \overline{\psi}{}^a_2\psi_1^b\overline{\psi}{}^c_2\psi_1^d \right]+\cdots\right)e^{-\int_0^1 ds \frac{1}{2}\left[\frac{1}{\sqrt{\Lambda}}J^r_3 \pm K^r_3 \right]\Delta \varphi} \right|_{J,K=0}
\nonumber \\
    =& \left.\left(1+\frac{T}{8\Lambda} [\Delta \varphi]^2\int_0^1 ds \delta_i^r \delta_j^r (\delta_{cg}\delta_{dh}-\delta_{ch}\delta_{dg})\frac{\delta}{\delta K^i_g}\frac{\delta}{\delta K^j_h} \left[ e^{-2i\vartheta} \overline{\psi}{}^3_2\psi_1^b\overline{\psi}{}_2^3\psi_1^d \right]+\cdots\right)e^{-\int_0^1 ds \frac{1}{2}\left[\frac{1}{\sqrt{\Lambda}}J^r_3 \pm K^r_3 \right]\Delta \varphi} \right|_{J,K=0}.
\end{align}
The second term in parentheses vanishes due to an anti-symmetry argument. Thus, the propagator expressed with the gauge fields $A^\pm$ is given by:
\begin{align}\label{eq20.17}
    & \int_0^{2\pi}\frac{d\vartheta}{2\pi}e^{e^{-i\vartheta} \overline{\psi}{}^a_2\psi^b_1(\delta_{ab}\mp \frac{1}{4} \epsilon_{3ab}[x^r_1-x^r_0]\Delta \varphi)-i(s-2)\vartheta} \underbrace{\int_0^\infty dT \int_{x^r(0)=x^r}^{x^r(1)=y^r} \mathcal{D}x \sqrt{\det [\varphi(z_1)-\varphi(z_0)]^2} e^{-\int_0^1 ds \frac{1}{16\Lambda T} [\varphi(z_1)-\varphi(z_0)]^2 (\dot{x}{}^r)^2}}_{\leadsto 1}
\nonumber \\
    =& \int_0^{2\pi}\frac{d\vartheta}{2\pi}e^{e^{-i\vartheta} \overline{\psi}{}^a_2\psi^b_1(\delta_{ab}\mp \frac{1}{4} \epsilon_{3ab}[x^r_1-x^r_0]\Delta \varphi)-i(s-2)\vartheta}
\nonumber \\
    =& \int_0^{2\pi}\frac{d\vartheta}{2\pi} \langle \psi_2 | e^{i\vartheta \left[\overline{\psi}{}^a \psi^b(\delta_{ab}\mp \frac{1}{4} \epsilon_{3ab}[x^r_1-x^r_0]\Delta \varphi)+\frac{5}{2}\right]} | \psi_1 \rangle.
\end{align}
In the first line, the term analogous to (\ref{eq19}) gives $1$ because the mass is zero for a photon, so introducing a small mass $\mu$ and defining the integration measure so that it compensates $\frac{1}{2\mu}$ gives the desired result. Now, upon writing compactly the expectation value of the propagator we obtain:
\begin{equation}\label{eq20.18}
    \left\langle \frac{1}{\gamma_{ij} \nabla_k \nabla^k-R^{\text{3D}}_{ij}} \delta^{(3)}(x,y)\right\rangle_\text{K.} = \left\langle \int_0^{2\pi}\frac{d\vartheta}{2\pi} \langle 0 | [\Delta \varphi]^2\psi^3 e^{i\vartheta \left[\overline{\psi}{}^a \psi^b(\delta_{ab}-(\pm 1) \times \frac{1}{4} \epsilon_{3ab}\Delta x^r \Delta \varphi)+\frac{5}{2}\right]} \overline{\psi}{}^3 | 0 \rangle \right\rangle_{\text{L.}^+ \times \text{L.}^-}.
\end{equation}
This is a compact notation and inside of the integral, we have to consider both signs in front of the term arising from the spin connection. Specifically, to have the full expectation value expression, we must replace $(\pm 1) \times \frac{1}{4} \epsilon_{3ab}\Delta x^r \Delta \varphi \leadsto \frac{1}{4} \epsilon_{3ab}\Delta x^r \Delta \varphi^+-\frac{1}{4} \epsilon_{3ab}\Delta x^r \Delta \varphi^-$, where $\varphi^\pm$ belongs to the expectation value on the Liouville model $\text{L.}^\pm$. To simplify further this expression (\ref{eq20.18}), we write $[\Delta \varphi]^2 = \left.-\partial^2_\eta e^{i\eta [\Delta \varphi]} \right|_{\eta =0}$. We are thus interested in the following expression:
\begin{align}\label{eq20.19}
    & \left\langle \int_0^{2\pi}\frac{d\vartheta}{2\pi} \langle 0 | \psi^3 e^{\left(-(\pm1)\times\frac{i}{4}\vartheta \overline{\psi}{}^a \psi^b  \epsilon_{3ab}\Delta x^r +i\eta \right) \Delta \varphi +\frac{5i\vartheta}{2}} \overline{\psi}{}^3 | 0 \rangle \right\rangle_{\text{L.}^+ \times \text{L.}^-}
\nonumber \\
    \leadsto& \int_0^{2\pi}\frac{d\vartheta}{2\pi} \langle 0 | \psi^3 e^{\frac{5i\vartheta}{2}} 2^{-2\left( \eta  +\frac{1}{16}\vartheta^2 \overline{\psi}{}^a \psi^b \overline{\psi}{}^c \psi^d  \epsilon_{3ab}\epsilon_{3cd}[\Delta x^r]^2\right)}\overline{\psi}{}^3 | 0 \rangle.
\end{align}
Where $\leadsto$ means we divide by the free partition function (the expectation value of 1). We can expand again into a series, and we obtain a term of the form $\epsilon_{3ab}\epsilon_{3cd} \psi^3\overline{\psi}{}^a \psi^b \overline{\psi}{}^c \psi^d \overline{\psi}{}^3 |0 \rangle \leadsto \epsilon_{3ab}\epsilon_{3cd} \delta^{33} \delta^{ad}\delta^{bc} |0 \rangle = 2|0 \rangle$, at the order 1. So finally, the expectation value (\ref{eq20.18}) is (we pose $|\Delta z|=R\tan\left( \frac{\theta}{2} \right)$):
\begin{align}\label{eq20.20}
    (\ref{eq20.18}) \sim& 4\ln(|\Delta z|)\int_0^{2\pi}\frac{d\vartheta}{2\pi}\left( -1+\frac{1}{4}\ln(|\Delta z|)\vartheta^2 (\Delta x^r)^2\right) e^{\frac{7i\vartheta}{2}} |\Delta z|^{-\frac{1}{4}\vartheta^2[\Delta x^r]^2}
\nonumber \\
    =& \lim_{\theta \rightarrow 0}4\ln\left( R\tan\left( \frac{\theta}{2} \right) \right)\int_0^{2\pi} \frac{d\vartheta}{2\pi}\left( -1+\frac{1}{4}\ln\left( R\tan\left( \frac{\theta}{2} \right) \right)\vartheta^2 (\Delta x^r)^2\right)
\nonumber \\    
    & \times e^{\frac{7i\vartheta}{2}} \left( \frac{4}{\theta^2 R^2}\right)^{\frac{1}{8}(\Delta x^r)^2 \vartheta^2}\left[ 1 - \frac{1}{48}(\Delta x^r)^2 \theta^2 \vartheta^2+\mathcal{O}(\theta^4)\right].
\end{align}
Where the right-hand side is understood as being the $i,j=r$ component of (\ref{eq20.18}). For $i,j \neq r$, the propagator becomes:
\begin{align}\label{eq20.21}
    (\ref{eq20.18})|_{i,j\neq r} \sim& \int_0^{2\pi} \frac{d\vartheta}{2\pi} e^{\frac{7i\vartheta}{2}} \left( \frac{4}{\theta^2 R^2}\right)^{\frac{1}{8}(\Delta x^r)^2 \vartheta^2}\left[ 1 - \frac{1}{48}(\Delta x^r)^2 \theta^2 \vartheta^2+\mathcal{O}(\theta^4)\right].
\end{align}
The propagator is zero if $i=r$ or $j=r$. Except for the prefactor and the integral, this is roughly the same expression as in \cite{Roux:2024bna}, with a mass squared $m^2=\frac{1}{2}\Lambda \vartheta^2$. Because the mass is directly proportional to the constant U(1) gauge field (which has little to do with the fact that we are describing a photon) which is gauge-dependent, we can say this mass is fictive.

\end{document}